\documentclass[journal]{IEEEtran}
\usepackage{cite}
\usepackage{amssymb,amsfonts}
\usepackage{algorithmic}
\usepackage{graphicx}
\usepackage{textcomp}
\def\BibTeX{{\rm B\kern-.05em{\sc i\kern-.025em b}\kern-.08em
    T\kern-.1667em\lower.7ex\hbox{E}\kern-.125emX}}
\usepackage[caption=false,font=normalsize,labelfont=sf,textfont=sf]{subfig}
\usepackage{graphics} % for pdf, bitmapped graphics files
\usepackage{epsfig} % for postscript graphics files
\usepackage{xcolor}
\usepackage[utf8]{inputenc}% http://ctan.org/pkg/inputenc
\usepackage{amsmath}% http://ctan.org/pkg/amsmath
\usepackage{siunitx}
\usepackage{adjustbox}
\usepackage{multirow}
\usepackage{xfrac}
\usepackage{tablefootnote}
\usepackage[hidelinks]{hyperref}
\DeclareUnicodeCharacter{2061}{}
\begin{document}

\title{Schr\"odinger Spectrum based Continuous Cuff-less Blood Pressure Estimation using Clinically Relevant Features from PPG Signal and its Second Derivative}

\author{Aayushman Ghosh$^\mathcal{y}$, Sayan Sarkar$^\mathcal{y}$, and Jayant Kalra

\thanks{$^\mathcal{y}$Both authors contributed equally. \\ This work of Aayushman Ghosh and Sayan Sarkar is partly supported by EUREKA start-up grant of IIT Bombay. The work of Jayant Kalra is supported by Industrial Research and Consultancy Centre, Indian Institute of Technology (IIT), Bombay. (Corresponding author: Sayan Sarkar)}
\thanks{Sayan Sarkar is associated with the Department of Electronic and Computer Engineering, The Hong Kong University of Science and Technology, Hong Kong, China (email: \href{mailto:ssarkar@connect.ust.hk}{ssarkar@connect.ust.hk}).}
\thanks{Aayushman Ghosh is associated with the Department of Electronics and Telecommunication Engineering, Indian Institute of Engineering Science and Technology (IIEST), Shibpur, Howrah 711103, India. (e-mail: \href{mailto:510719076.aayushman@students.iiests.ac.in}{510719076.aayushman@students.iiests.ac.in}).}
\thanks{Jayant Kalra is associated with the Department of Metallurgical Engineering and Materials Science, Indian Institute of Technology (IIT), Bombay, Mumbai 400076, India. (e-mail: \href{mailto:iamjayantkalra@gmail.com}{iamjayantkalra@gmail.com}).}}

\markboth{\textbf{PREPRINT} \\ 
SUBMITTED TO BIOMEDICAL SIGNAL PROCESSING AND CONTROL, ELSEVIER}%
{Shell \MakeLowercase{\textit{et al.}}: Bare Demo of IEEEtran.cls for IEEE Journals}

\maketitle
%%%%%%%%%%%%%%%%%%%%%%%%%%%%%%%%%%%%%%%%%%%%%%%%%%%%%%%%%%%%%%%%%%%%
%\vspace{-15mm}
\begin{abstract}

The presented study aims to estimate blood pressure (BP) using photoplethysmogram (PPG) signals while employing multiple machine-learning models. The study proposes a novel algorithm for signal reconstruction, which utilises the semi-classical signal analysis (SCSA) technique. The proposed algorithm optimises the semi-classical constant and eliminates the trade-off between complexity and accuracy in reconstruction. The reconstructed signals' spectral features are extracted and incorporated with clinically relevant PPG and its second derivative’s (SDPPG) morphological features. The developed method was assessed using a publicly available virtual in-silico dataset with more than 4000 subjects, and the Multi-Parameter Intelligent Monitoring in Intensive Care Units dataset. Results showed that the method attained a mean absolute error of 5.37 and 2.96 mmHg for systolic and diastolic BP, respectively, using the CatBoost supervisory algorithm. This approach met the standards set by the Advancement of Medical Instrumentation, and achieved Grade A for all BP categories in the British Hypertension Society protocol. The proposed framework performs well even when applied to a combined database of the MIMIC-III and the Queensland dataset. This study also evaluates the proposed method’s performance in a non-clinical setting with noisy and deformed PPG signals, to validate the efficacy of the SCSA method. The noise stress tests showed that the algorithm maintained its key feature detection, signal reconstruction capability, and estimation accuracy up to a 10 dB SNR ratio. It is believed that the proposed cuff-less BP estimation technique has the potential to perform well on resource-constrained settings due to its straightforward implementation approach. 

\end{abstract}


\IEEEpeerreviewmaketitle

\section{INTRODUCTION}
\IEEEPARstart{C}{ARDIOVASCULAR DISEASES} (CVDs) are a leading cause of morbidity and mortality worldwide. Common CVDs include coronary heart disease, hypertension and stroke [1–4]. Physicians widely adopted different techniques to monitor an individual's cardiovascular health to prevent CVDs effectively. One such technique is the periodic measurement of blood pressure (BP). Hypertension is one of the prevailing CVDs worldwide. It shows a significant risk factor that can damage vital organs like the heart, brain, and kidneys [5]. While medications can manage hypertension (high BP) but BP can fluctuate rapidly over time and is influenced by factors such as stress, emotions, food, exercise and medications. So, continuous and accurate BP measurement becomes crucial in managing a person’s overall health [2–5].

Medical practitioners either use non-invasive cuff-based techniques, such as auscultation, volume–clamp, and applanation tonometry or rely on invasive procedures, such as arterial lines [2–10], for BP measurement. Cuff-based methods provide intermittent measurements, which are unreliable for continuous monitoring over long periods and result in discomfort and inconvenience for the patients. Arterial lines, while accurate and continuous, require catheterisation and strict medical intervention [1–12]. Non-invasive cuff-less methods, which use physiological signals to estimate BP, can be a viable alternative to traditional cuff-based techniques [1, 2, 5–10]. These methods have gained significant attention in recent years and can be divided into two categories: (1) those that use only photoplethysmography (PPG) signals [4, 16] and (2) those that use both PPG and electrocardiography (ECG) signals [11, 15]. The first category involves using one or more PPG signals (either the entire signal or extracted features) to estimate BP, while the second category combines both PPG and ECG signals to estimate BP. In this study, the authors have mainly focused on techniques that only use PPG signals, with a brief overview of methods using PPG and ECG signals, as shown in Fig. 1 and in section 2.1. The PPG signal can be thought of as a low-pass filtered version of the Arterial Blood Pressure (ABP) waveform that provides valuable information about the cardiorespiratory activity [3–6] (refer Fig. 1). Each PPG signal has distinct anacrotic and catacrotic phases corresponding to the rising and falling curve of the PPG, which varies between systolic and diastolic Blood Pressure in each heartbeat interval. The systolic part of the PPG signal is associated with cardiac contraction, while the diastolic part is related to cardiac expansion [3–14]. The duration of the systolic and diastolic phases of the PPG pulse wave can affect its shape in an individual. However, various physiological processes (such as respiration), circulation factors (such as arterial stiffness and BP), cardiorespiratory aspects (like the heart's rhythmic movement and stroke volume), and diseases can also alter the shape of the PPG signal.

Several studies have investigated cuff-less computational strategies to estimate BP using only PPG signals [6–20]. These approaches can be majorly distinguished into two procedures: (1) Pulse Wave Velocity (PWV), and (2) Pulse Wave Analysis (PWA). PWV (strong correlation with BP) techniques estimate BP by computing the propagation velocity of the pressure wave [15]. Some of these techniques measure the Pulse Transit Time (PTT) [12], which is the time taken by the PPG pressure wave to travel between two arterial points and is inversely proportional to the PWV (Fig. 1). Other techniques calculate the Pulse Arrival Time (PAT), which is the interval between the ECG R-peak and a characteristic point on the PPG signal [11]. One of the major challenges with using PWV techniques is that they may not be consistently accurate at different locations in the body due to the lack of elasticity in the terminal arteries compared to the central arteries [14–16]. These techniques also require additional sensors to capture PPG at different arterial locations, which limits their practicality and usability [3–6]. In contrast, PWA methods rely solely on single PPG signals and involve analysing the shape of the PPG pressure waveform to extract features that can be used to estimate BP. Several time-based, amplitude-based, and frequency-based features have been proposed in earlier literature for this purpose [7–27]. The components extracted from the PPG pressure waveform are generally mapped to BP using mathematical algorithms [22] or various machine learning (ML) algorithms such as multiple linear regression (MLR) [20], Artificial Neural Networks (ANN) [21], Support Vector Machine (SVM) [19], Random Forest [11], etc. PWA typically requires an initial calibration using a brachial cuff to track BP. This method accurately estimates relative changes in BP rather than absolute values. PWA analysis faces challenge when the quality of the acquired PPG signal deteriorates [16–22], [31–33] in watch or smartphone-based PPG signal acquisition. The lack of significant peaks and ample amount of noise further worsen pressure waveform processing. This can significantly reduce the accuracy of the BP measurement.

\begin{figure*}
     \centering
     \includegraphics[width = 18.1cm, height = 5.0cm]{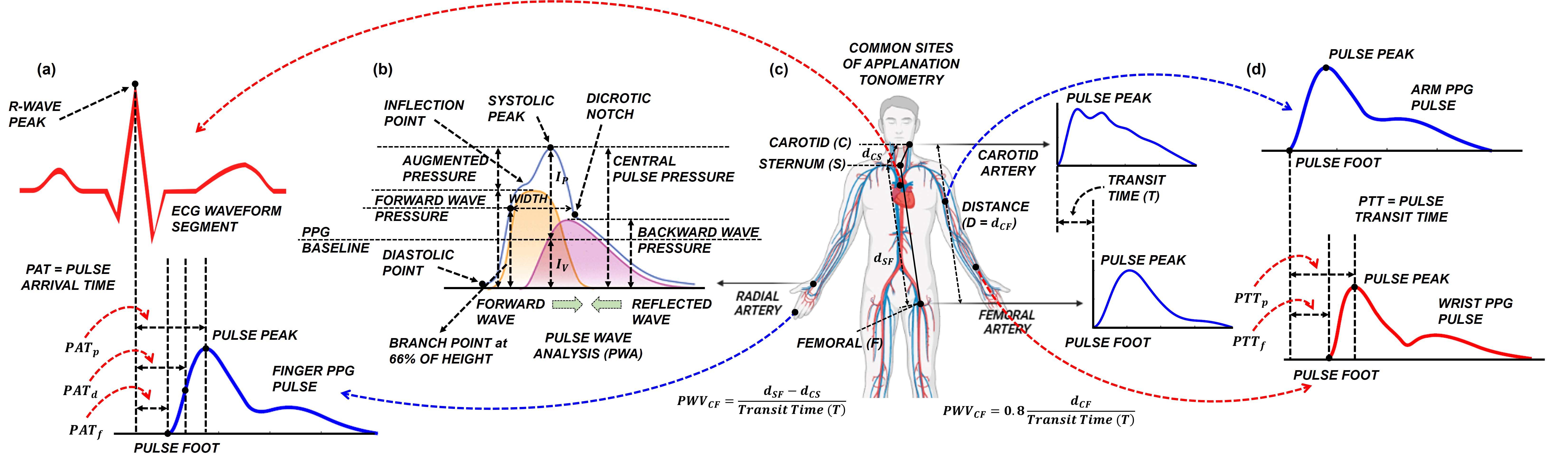}
     \caption{(a) Typical representation of ECG and PPG wave with different PAT features. (b) Time domain features in PPG showing with PDA principle – standard backward and forward pulse-based decomposition. (c) Arterial wave propagation model in the human body with common sites of applanation tonometry. (d) Typical representation of a time-shifted pulse wave with different types of PTT features.}
     \label{figure-1}
\end{figure*}

A new approach, Pulse Decomposition Analysis (PDA), was proposed to tackle the issue faced by PWA techniques [28–30]. It is based on the idea that in addition to the primary systolic pulse, two additional pulses are reflected from the renal and iliac arteries branching off the abdominal aorta [28]. PDA relies on analysing these reflected waves. The magnitude and location of these reflected pulses are influenced by BP and flow velocity, and by fitting this information into a model it is possible to estimate changes in BP [29]. This technique can be instrumental when there is inconsistency in the shape of the PPG signal. The underlying theory of PDA algorithms can be mathematically expanded further by decomposing the parent wave (here the PPG signal) into multiple sub-waves (or reflected waves). The Semi-Classical Signal Analysis (SCSA) is one such approach that extends the capabilities of PDA algorithms by using a mathematical decomposition method called the non-linear superposition of solitary waves or solitons [48–53]. This method involves using solitons, which are solutions of non-linear partial differential equations, to analyse pulse-like signals such as PPG. In this approach, the PPG is typically represented as a superposition of multiple solitons, and the soliton with the highest amplitude is considered to be the most significant [48–51]. These solitons are further used to reconstruct the original PPG signal. SCSA has been used to analyse various pulse-like signals, including those related to BP measurement. Laleg \textit{et al.} [51, 52] attempted to use the SCSA algorithm on PPG signals. Still, the strategy did not result in superior clinical accuracy, thereby failing to achieve optimised parameters for reconstruction [48–53]. Moreover, the generalisation capability of Laleg’s strategy faces serious questions as their study lacks promising analysis in different BP categories. To solve the existing problem, authors have proposed a new error feedback-based reconstruction algorithm to optimise the reconstruction parameters, which allows for a balance between the number of solitons and the computational burden. This modified approach to the original SCSA algorithm has improved signal reconstruction. Here authors have further combined PWA-based features from PPG and SDPPG with spectral features obtained through SCSA to achieve clinical accuracy. These combined strategies significantly improve BP estimation accuracy. The study has also been extended to multiple public databases to demonstrate the generalizability of the proposed algorithm. As per the authors' best belief, this is the first study to incorporate in-silico [55], multiple publicly available [31], and wearable-collected noisy datasets [65] for the evaluation of a BP algorithm. The main contributions of the study are summarised as follows:
\begin{enumerate}
    \item  Introduction of a modified SCSA framework that optimises the $h$-value associated with signal reconstruction to achieve superior reconstruction performance with a minimised computational load.
    \item Generalising the capability of the proposed algorithm by checking its robustness against (a) noisy wearable collected; and (b) artificial noise (different SNRs) induced datasets. 
    \item Evaluating the effectiveness of the proposed framework in different BP categories (hypertension/normotension/ hypotension) using clinically acquired datasets.
    \item Introducing clinically relevant PPG and SDPPG-derived features along with SCSA–based spectral features to enhance the BP estimation method's performance and obtain the advantages of both PWA and PWD techniques in noisy PPG signal.
\end{enumerate}

The structure of the paper is as follows: Section 2 provides a brief overview of traditional approaches and recent developments in BP estimation methodology. Section 3 describes the dataset used in the study, the SCSA algorithm, the extracted features’ description, and the machine learning models used. Section 4 presents the results and discussions of the proposed methods, a comparison with state-of-the-art approaches, and a critical analysis using multiple publicly available datasets. Finally, Section 5 concludes the paper with a summary of the results and critical analysis.

\section {REVIEW OF EARLIER LITERATURES}
Blood pressure (BP) is a vital sign that measures the force of blood pumping through the circulatory system [1]. Traditional cuff-based techniques, commonly used in clinical and non-clinical settings,  may not be suitable for long-term use. An alternative method for measuring BP is the cuff-less method, especially using photoplethysmography, which extracts several parameters that can be used to estimate BP. These parameters, known as BP measurement modalities, include PWV and PWA techniques. These approaches have been well-researched in previous literature [6–27], but there is limited information available on PDA. 

\vspace{-2.5mm}

\subsection {Transit Time Calculations and PWV}
The study conducted by Teng \textit{et al.} was the first to introduce a PPG-based BP evaluation framework. They extracted standalone features from PPG and adopted a linear regression approach for BP estimation [7]. McCombie \textit{et al.} [13] suggested that the PWV-based BP estimation strategy has difficulties as the peripheral arterioles lack the central arteries' elasticity [3]. PWV is the pressure pulse’s velocity through the arterial system and is typically calculated using transit time measurements such as pulse transit time (PTT), which is the time taken by a pressure pulse to travel from one arterial location to another [11–16]. However, this approach may be inadequate as it ignores the variations in arterial diameter and relies on a complex model of arterial wave propagation [14–16]. These factors introduce errors during the calculation of BP. The arterial elasticity is subject-specific and substantially relies on age, height, weight, diet, arterial length, etc., necessitating periodic calibration [14]. Recent studies [11, 20, 25, 27] indicate a shift from PTT to PAT-based estimation for better accuracy. PAT is substantially influenced by the cardiac pre-ejection period (PEP) before the aortic valve opening [14]. PEP contains the time required to transform the electrical signal into a mechanical pumping force and the left ventricle's isovolumetric contraction to open the aortic valve [14]. Kachuee \textit{et al.} [11] suggested using PAT, or pulse arrival time, as a feature to help estimate BP using multiple machine learning frameworks. Our previous research [15] also utilised PAT as a single feature in estimating BP using gradient-boosting algorithms. While PAT can lower the accuracy of estimating diastolic BP [27], it is commonly used due to its simplicity [11–12]. The relationship between PWV (pulse wave velocity), PTT (pulse transit time), and PAT, as well as their definitions, are shown in Fig. 1.

\vspace{-2 mm}

\begin{figure*}
     \centering
     \includegraphics[width = 17.7cm, height = 7.5cm]{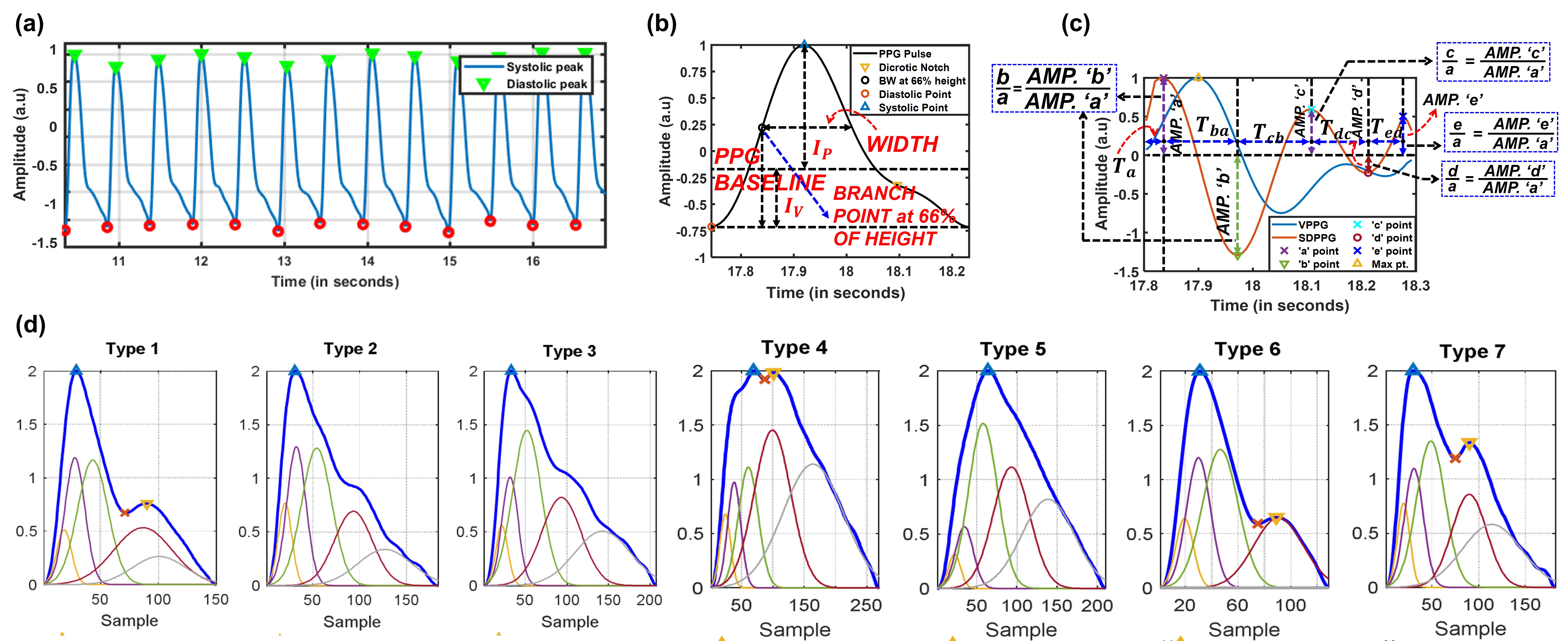}
     \caption{(a) A sample of implementation of the Automatic Multiscale-based Peak Detection (AMPD) algorithm in detecting the systolic peak and diastolic point of a PPG signal (b) Location of time-domain and clinically relevant features of a PPG pulse and (c) Location of peak points, amplitude-ratio and timestamps of a VPPG and SDPPG pulse. (d) Morphology of PPG signal – as a decomposition of 5 component Gaussian waves (g1, g2, g3, g4, g5) in related subclasses, discriminated on behalf of their shape – type 1, type 2, type 3, type 4, type 5, type 6, and type 7. Panel (d) is reprinted with the permission of [45], © The Authors, IEEE.}
     \label{figure-2}
\end{figure*}

\subsection{Pulse Wave Analysis of PPG}
PPG-based PWA [16, 17] is a standard method for evaluating arterial stiffness, BP, and understanding the progression of diseases [3]. It involves analysing the shape of the PPG pulse waveform to extract features, which are then used with ML algorithms [16–27] or neural networks to estimate or classify one of the above. In this study, the authors focused only on BP estimation. Initial studies considered multiple PWA parameters such as systolic upstroke time, diastolic time, width of $\sfrac{2}{3}$ or $\sfrac{1}{2}$ of the pulse amplitude [7] as potential BP indicators from a single PPG pulse. Kurylyak \textit{et al.} [8] extracted similar features from PPG waveforms and developed a feed-forward neural network for SBP and DBP estimation. Khalid \textit{et al.} [20] also studied the use of pulse area, $25\%$ width, and pulse rising time as PPG features for BP estimation using multiple linear regression (MLR), support vector regression (SVR), and decision tree regression (DTR). They found that the decision tree algorithm provided the most accurate estimation, but their analysis was limited to a small sample size of $32$ subjects from the Queensland dataset, and the results were unsatisfactory. Ding \textit{et al.} [18] on the other hand, introduced a new feature, the Photoplethysmogram Intensity Ratio, $(PIR_p = \sfrac{I_p}{I_v})$ for BP estimation (see Fig. 2(b) for definition). Recent research directions also involve studying the second derivative of the PPG signal (SDPPG). The SDPPG signal, which has a `W-shaped' appearance, has been shown to contain information about arterial compliance and aortic stiffness [3]. The relationship between vascular ageing and the SDPPG signal was first suggested in [20]. Five characteristic points (`$a$', `$b$', `$c$', `$d$', and `$e$') on the SDPPG signal (which are identified as peaks and troughs) and their corresponding timestamps ($T_a$, $T_{ba}$, $T_{cb}$, $T_{dc}$, $T_{ed}$) are frequently used in current research to provide clinically relevant insights [3]. For instance, changes in the amplitude ratios of $\sfrac{c}{a}$, $\sfrac{d}{a}$, and $\sfrac{e}{a}$, as well as an increase in $\sfrac{b}{a}$, are thought to indicate increased arterial stiffness [3]. The $\sfrac{b}{a}$ and $\sfrac{c}{a}$ ratios are frequently used to distinguish between hypertensive and healthy subjects. Liu \textit{et al.} [19] extended Kurylyak's study by adding $14$ more features from the SDPPG signal. Duan \textit{et al.} [24] examined $57$ possible features (divided into three sets of $11$ features) to predict BP using an SVR algorithm. The accuracy of their results barely met the clinical standards set by the Association for the Advancement of Medical Instrumentation (AAMI) and the British Hypertension Society (BHS). In a study by Yi \textit{et al.} [25], various algorithms (linear regression, elastic network, LASSO, KNN, and CART) were applied to $9$ morphological features extracted from PPG signals to estimate BP. The results showed that the K-nearest neighbours (KNN) algorithm performed the best and that using a small number of relevant PPG features could achieve accuracy similar to using multiple PPG features. The study concluded the necessity of identifying relevant features that are effective across a wide range of datasets in standardising the PWA method [16, 31]. Some features, such as $BW_{66}$ and $PIR_p$ emerged, as they provide a strong clinical correlation with BP. $BW_{66}$ is related to the pulse width of the PPG signal [31], and $PIR_p$ has been previously discussed [18] (see Fig. 2(a) for definition). In a recent study, the authors have identified clinically relevant features for estimating BP, resulting in $ME\pm SD$ values of $\sim0.15\pm6.24$ mmHg and $\sim0.06\pm3.51$ mmHg for systolic (SBP) and diastolic (DBP) blood pressure, respectively. However, this study was limited to a small subset of the MIMIC-II database $(N=90)$ [34]. While PPG features are commonly used, they are pre-defined and generic [16–19], and extracting relevant features from PPG signals can be challenging due to variations in the morphology of the signal, particularly in the elderly and in the presence of noise. Therefore, some studies have used basic deep learning architectures, such as artificial neural networks (ANN), convolutional neural networks (CNN), and long short-term memory (LSTM) networks, to analyse the entire PPG signal rather than extracting specific features [35, 37, 43]. A recent study suggests that including subject-specific details in the raw PPG signal could improve the accuracy of BP estimation [16]. Wang \textit{et al.} [23] combined both NN-based estimation and feature extraction in their study. They separated single PPG pulse strands from raw signals in his study and then extracted their morphological and spectral features using a wrapper method. These features were then used to train an ANN for BP prediction. While Deep Learning frameworks can provide improved accuracy, they require a large number of parameters and can significantly increase the computational burden of the system [35].

\vspace{-2 mm}
\subsection{Pulse Decomposition of PPG}
PDA is a technique used to measure hemodynamic parameters, such as blood pressure, by analysing pulse contours. PDA is based on the idea that the shape of the pressure pulse envelope in the upper body is primarily influenced by central reflection sites rather than distal sites in the arterial periphery. According to PDA, the observed pulse shape comprises five individual component pulses. ‘Left ventricular ejection pulse’ is the first of these component pulses to reach the arterial periphery, followed by reflections and re-reflections from two central arteries' reflection sites. PDA posits that quantifying these components' temporal and amplitudinal behaviour allows the monitoring of specific hemodynamic parameters and their changes [28, 29, 38, 39, 44, 45].

The choice of basis functions in decomposition is essential because it can introduce a quantitative bias to the results and impact the reconstruction performance [28–30]. The most effective basis function is the Gaussian wave function [38, 39]. For example, in [45], the authors used a weighted pulse decomposition approach with the Gaussian function because it has a similar shape to pulsatile waves. They also divided the PPG into seven types based on their general shape and the artery location from which they were obtained (see Fig. 2(c)). Other models that utilise different basis waves, such as the Rayleigh, log-normal, and Gamma functions, have also been examined for pulse fitting [39]. As an alternative to these popular basis functions, the use of secant hyperbolic waves (Sech) for the PDA of PPG signals was considered in [44]. The `Sech' function may provide a better PDA for blood pressure estimation because it is a potential solution to the Moens-Korteweg equation, which describes a pressure pulse in an elastic tube. In [44], the results suggested that the hyperbolic secant wave function performs better for three-wave PDA than previous functions in the PDA literature [28–30, 38], as `Sech' reduces pulse-to-pulse parameter noise in the reflected waves. However, these PDA techniques have a common disadvantage: the basis functions are well-defined mathematical relations that only partially fit a proper PPG signal. The reconstruction loss [45] using these well-defined basis functions becomes significant when the morphology of the PPG signal changes. Laleg \textit{et al.} [48] proposed the Semi-Classical Signal Analysis (SCSA) Theory to address these limitations. Instead of the standard linear decomposition strategy [38, 44, 45], the SCSA approach uses a non-linear superposition of solitary waves or solitons (derived orthogonal basis functions) as the decomposition strategy [46]. The PPG signal is typically shown as a superposition of `$N$' solitons with the soliton with the largest amplitude being the most significant. The remaining solitons characterise the noise details of the PPG waves [46, 49]. The sum of the first $2$ or $3$ solitons represents the fast phenomena of the PPG signal, the systolic phase. Generally, a two-element windkessel model describes the slow phenomena during the diastolic phase [46]. Alfonso \textit{et al.} [47] modelled the BP waveform as a combination of solitons, separating BP's fast and slow dynamics that correspond to the systolic and diastolic phases, respectively. In the SCSA approach, the PPG signal is considered a potential of a Schr\"odinger operator, which is then decomposed into a set of negative eigenvalues and corresponding squared eigenfunctions. Each generated negative eigenvalue is related to one soliton in terms of their velocities. However, the estimation problem arises because the number of components needed for an accurate soliton representation of the PPG signal is not a priori known, along with their respective velocities and widths [48–50]. This makes the standard SCSA algorithm a burdensome and complicated approach.

In this paper, the authors aim to address this issue of estimating the no. of components and their respective velocities by introducing a new algorithm (error-feedback-based reconstruction; error = reconstructed – input) into the SCSA framework. The authors are interested in achieving excellent reconstruction performance and estimating continuous blood pressure by collecting features from the reconstructed PPG signal and mapping them using supervised ML algorithms.

\vspace{-2 mm}
\section{MATERIALS AND METHODS}
This section will describe the dataset used in this study, the pre-processing step, the proposed SCSA methodology, and the machine learning (regression) algorithms and strategies we used to estimate continuous BP. 

\vspace{-2mm}
\subsection{Database Description}
For this study, the authors have primarily used the cuffless BP dataset from the Machine Learning Repository of the University of California, Irvine (UCI) to train and test the developed framework. The dataset contains 1200 signal segments of recorded ECG, PPG, and ABP signals from nearly 1000 patients [16]. The source of this dataset was the Physionet's MIMIC-II database [54], which was pre-processed and placed in this repository by Kachuee [11]. The characteristics of the used dataset are well described in previous literature [11,16]. Each data file in the UCI dataset includes three bio-signals: (a) lead II ECG, (b) fingertip PPG, and (c) ABP signal. Based on an observation of the clinical setup in the intensive care unit (ICU) and consultation with physicians, a data exclusion and selection process was applied [15]. First, the ECG recording from each data file was removed as the analysis only involved PPG signals. A check was then applied to see if the PPG and ABP signal length was sufficiently long ($\geq10$ min). Second, the original signal span's first and last $30$ seconds were removed as motion artefacts often affect them [15, 34]. Finally, a randomly selected $100$-second window was slid on the waveforms to extract the final PPG and ABP waveforms. The UCI dataset is not outlier-removed and, therefore, cannot be used directly in machine learning models [11, 14–16]. The available waveforms contain segments where the bio-signals are substantially deformed due to saturation or unknown reasons generated within the ICU [53]; therefore, pre-processing is essential. Several exclusion criteria were applied to remove physiologically irrelevant BP values, such as filtering out ABP signals with very high and very low systolic and diastolic blood pressure values (systolic BP $\geq180$, diastolic BP $\geq130$, systolic BP $\leq80$, diastolic BP $\leq60$) [16].

\vspace{-2mm}
\subsection{Pre-Processing and Detection of Characteristic points}
Noise components infiltrated the raw signals, so the input PPG was passed through a 4th–order Chebyshev–II band-pass filter with a stop-band attenuation of 80 dB and upper and lower cutoff frequencies of $0.5 - 20$ $Hz$ [15, 34]. The algorithm uses zero-phase filtering to keep the group delay constant. The Chebyshev-II filter was chosen because it makes the dicrotic notch of the processed PPG signal more apparent than Butterworth filters [34]. PPG signals also contain a slowly varying DC component based on an individual's heart rate. It was removed by normalising the PPG waveform using the `$z$'-score value, i.e., by making the mean zero and the standard deviation equal to one. A 9th-order polynomial was fitted to the waveform to remove the low-frequency components [15]. The recordings in the UCI dataset contain waveforms with distorted shapes and distinct spikes that can affect the identification of characteristic points. As a result, the PPG signals are further passed through a $19$-sample window Hampel filter to remove this unwanted effect at the pre-processing step partially. The ABP signals were also passed through a $4^{th}$-order $19$-frame length Savitzky-Golay filter to preserve the systolic and diastolic peaks [15]. The pre-processing steps are illustrated in Fig. 4(a).

Before extracting features from the pre-processed (refined) PPG signal, it is essential to identify the characteristic points (refer to Fig. 2(b)) and divide the signal into pulses for further analysis. As PPG signals have different morphologies (refer to Fig. 2(c)), the point detection algorithm should be minimally sensitive to different morphologies. The study employs an Automatic Multiscale-based Peak Detection (AMPD) algorithm to detect the systolic and diastolic peaks of the PPG signal (refer to Fig. 2(a)) [16]. The algorithm has several advantages, the most attractive being its ability to withstand the short-term variation of the DC of the PPG signal [16]. After the detection of the systolic and diastolic points of the PPG signal, individual pulses lying between two consecutive diastolic points were extracted for further analysis. Each extracted pulse was then passed through a three-step procedure to extract the dicrotic notch. First, the time-series data (PPG signal) is subtracted from the straight line connecting the systolic and diastolic points, and the minimum is calculated. This point is marked as the dicrotic inflexion point. Second, a window of radius $\sfrac{RR}{5}$ ($RR$ is the median heartbeat interval computed using the diastolic points) is formed by taking the dicrotic inflexion point as the mean, capturing a part of the signal. Third, the second derivative of the captured signal (SDPPG) is calculated, and the minimum is designated as the dicrotic notch (refer to Fig. 2(b)). Post-detection of the characteristic points, a moving average algorithm with a length of five pulses was applied to remove the errors and missed detections associated with the detection of those points. Pre-processing and the subsequent key point detection were systematically performed using MATLAB 2020b (MathWorks, USA).

\begin{figure*}
    \centering
    \includegraphics[width=18.1cm, height = 8.5cm]{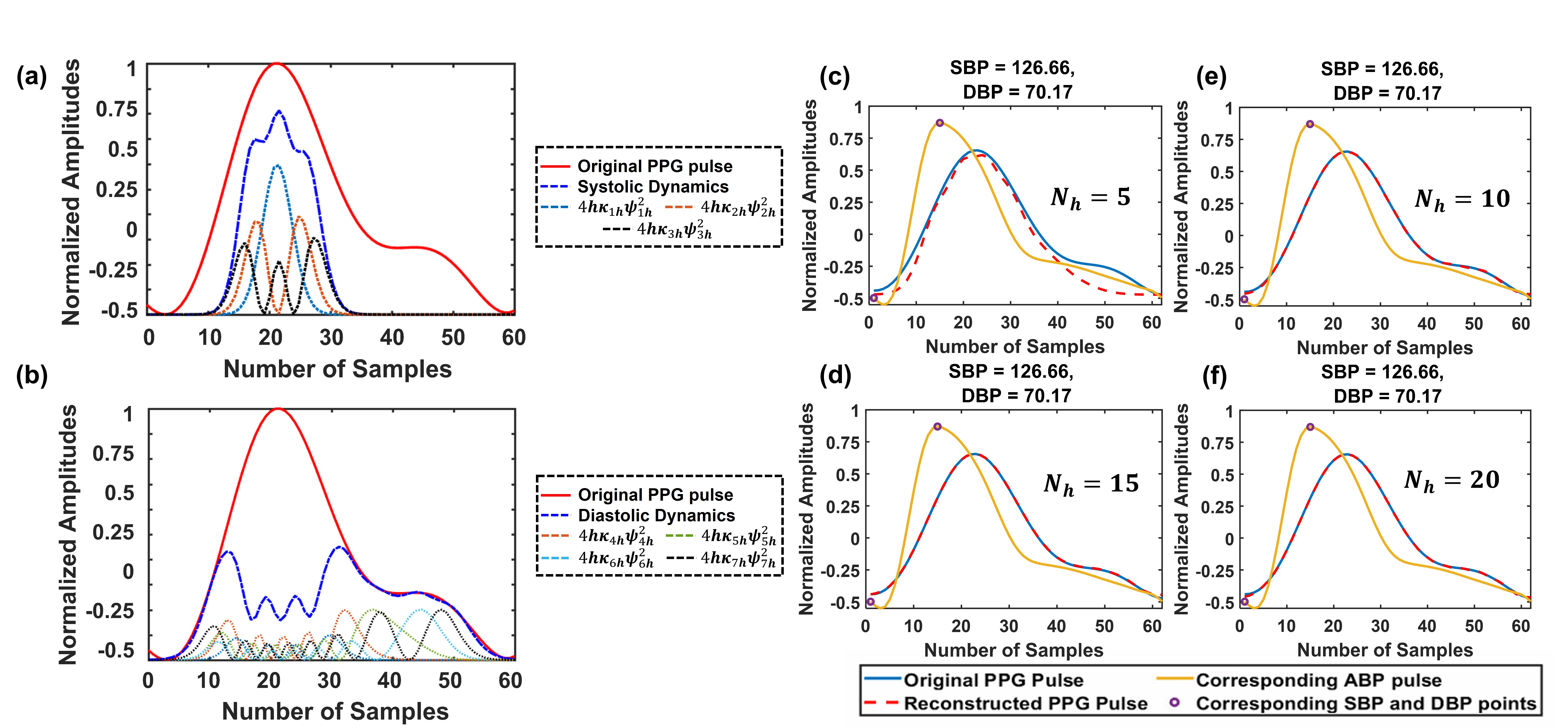}
    \caption{Signal decomposition and reconstruction using Semi-Classical Signal Analysis Theory. (a) Estimated Systolic Dynamics with PPG signal (in red) and SCSA estimated systolic part (in blue dotted line); (b) Estimated Diastolic Dynamics with PPG (in red), SCSA estimated diastolic part (in blue dotted lines) and Schrödinger components (colour coded according to the legend); Time lag between ABP (in yellow), PPG (in blue) and reconstructed PPG (in red dotted line). Pulse reconstruction using SCSA with different $N_h$ values (c) $N_h = 5$; (d) $N_h = 15$; (e) $N_h = 10$; (f) $N_h = 20$. As $N_h$ increases, perfect reconstruction is observed.}
    \label{figure-3}
\end{figure*}

\vspace{-2mm}
\subsection{Semi Classical Signal Analysis (SCSA) – Error-Feedback Based Reconstruction}

Laleg \textit{et al.} [48] in their seminal paper in 2010, proposed the theory of Semi Classical Signal Analysis as a signal decomposition and reconstruction strategy. The idea behind SCSA was initially inspired by the scattering transform (direct and indirect) based signal reconstruction [49]. The theoretical foundation for this idea is established by considering a real-valued non-negative function $y:t\xrightarrow[]{}y(t)$. Physically, $y$ is interpreted as the signal under test (in this case, the PPG pulse). In this study, the authors have interpreted the signal $y$ as a multiplication operator, $\phi \xrightarrow[]{}y.\phi$, defined on some function space. SCSA decomposes $y(t)$ into a set of squared eigenfunctions using the discrete spectrum of a Schrödinger operator. The Schr\"odinger operator (similar to the Hamiltonian in Quantum Mechanics) is therefore defined as (1).
\begin{equation}
    \mathcal{H}_h(t) = -h^2\frac{d^2}{dt^2} - y(t), \quad  h>0
\end{equation}
where the signal $y(t)$ is represented as the potential of the Schr\"odinger operator $\mathcal{H}_h$, where $h\in \mathbb{R}_{>0}$ is known as the semi-classical parameter. For this study, the authors are particularly interested in solving the associated spectral problem. In the general mathematical context, the spectral problem can be formulated as an eigenvalue problem of $\mathcal{H}_h(t)$ [50], where the Direct Scattering Transform (DST) method is used to find the solution. The eigenvalue problem follows (2).
\begin{equation}
    \mathcal{H}_h(t)\psi(t) = \lambda\psi(t), \quad  \psi \in H^2(\mathbb{R}); \mathcal{H}_h \in \mathcal{B}
\end{equation}
Where $\psi(t)$ is an eigenfunction on which $\mathcal{H}_h(t)$ operates, $\lambda$ is an eigenvalue, and $H^2(\mathbb{R})$ indicates the Sobolev space of order 2. Furthermore, $\mathcal{H}_h$ is defined on a space $\mathcal{B}$, where $L_1^1(\mathbb{R})$ is known as the Faddeev class [48–53].
\begin{equation}
\mathcal{B} \in\left\{\begin{array}{c}
y \in L_1^1(\mathbb{R}), y(t) \geq 0, \quad \forall t \in \mathbb{R} \\
\frac{\partial^m y}{\partial x^m} \in L^1(\mathbb{R}), \quad m=1,2
\end{array}\right.
\end{equation}
The solution of (2) is subject to boundary conditions, and because of that, the possible solution of eigenvalues is generally limited to a discrete set of eigenvalues $(\lambda<0)$ and a continuous set $(\lambda>0)$ over some range. The set of all such possible eigenvalues is referred to as the spectrum of (2). In the continuous spectrum, the asymptotic behaviour of the eigenfunctions is assumed, which is compactly written as 
\begin{equation}
    \psi(t,\kappa)\xrightarrow[]{}\mathcal{T}(\kappa)e^{-i\kappa x}, x\xrightarrow[]{}-\infty
\end{equation}
\begin{equation}
    \psi(t,\kappa)\xrightarrow[]{}e^{-i\kappa x} + \mathcal{R}(\kappa)e^{i\kappa x}, x\xrightarrow[]{}+\infty
\end{equation}
Where, $\mathcal{T}(\kappa)$ and $\mathcal{R}(\kappa)$ are the corresponding Transmission and Reflection coefficients. The other spectrum (discrete) is our point of interest, which will be clarified in the subsequent text. The discrete set of negative eigenvalues is denoted by $\lambda = -\kappa_{nh}^2$, $\kappa_{nh} > 0$, $n = 1, 2, 3,\cdots, N_h$, where $N_h$ denotes the total number of negative eigenvalues. In the discrete spectrum, the associated eigenfunctions $\psi_{nh}$ are $L^2$–normalised such that: $\int_{-\infty}^{+\infty}\psi_{nh}(t)^2 dt = 1$. The set $\mathcal{S} = \{\mathcal{R}(\kappa), \kappa_{nh}\}$ is called scattering data. Famously, this set is referred to as the solution provided by DST on $y$. The Indirect Scattering Transform (IST), on the other hand, uses $\mathcal{S}$ to reconstruct the potential of the Schr\"odinger operator, in this case, the PPG signal under test $y(t)$ [49]. Deift and Trubowitz [49] in their seminal work in the 1980s, established a generalized expression for the reconstructed function $y_h(t)$. They proposed that if $y(t)$ satisfies the three conditions, namely (a) $y$ must be infinitely differentiable to ensure it is smooth and continuous; (b) $y$ must asymptotically go to 0, according to $\int_{-\infty}^{+\infty}|y|(1+|t|)dt < \infty$; and (c) -$y$ must be attractive, then it can be expressed as
\begin{equation}
    y_h(t) = \frac{2i}{\pi}\int_{-\infty}^{+\infty}\mathcal{KR}(\kappa)\psi^2(t,\kappa)d\kappa + 4\sum_{n=1}^{N_h}\kappa_{nh}\psi^2_{nh}
\end{equation}
Where $\mathcal{K}=2\pi/t$, and $\psi(t,\kappa)$ is the solution corresponding to the continuous spectrum. The other terms of (6) have their usual significance. The computational burden of (6) becomes huge when iterating over multiple instances. A significant way of reducing that burden is to assume the signal $y(t)$ as a reflectionless potential, for which $\mathcal{R}(\kappa)=0$. This operation drives the left-hand side term of (6) to zero. Mathematically, now the signal can be reconstructed using only the discrete spectrum of eigenvalues, according to (7).
\vspace{-1mm}
\begin{equation}
    y_h(t) = 4\sum_{n=1}^{N_h}\kappa_{nh}\psi^2_{nh}(t)
\end{equation}
However, most of the bio-signals are not reflectionless. Nevertheless, Laleg \textit{et al.} [48] showed that if any bio-potentials satisfy the previously mentioned points (a)–(c), they can be assumed as reflectionless potential and can be reconstructed using the discrete spectrum of $\mathcal{H}_h(t)$ in the semi-classical limit of $h\xrightarrow{}0$. In conclusion, $h$ helps in approximating the input signal $y(t)$ using only the reflectionless part of the potential, as per (8).
\vspace{-1mm}
\begin{equation}
    y_h(t) = 4h\sum_{n=1}^{N_h}\kappa_{nh}\psi^2_{nh}(t), \quad t \in \mathbb{R}; h = \frac{1}{\chi^2}
\end{equation}
\begin{figure*}[ht]
    \centering
    \includegraphics[width = 18.1 cm, height = 8.3cm]{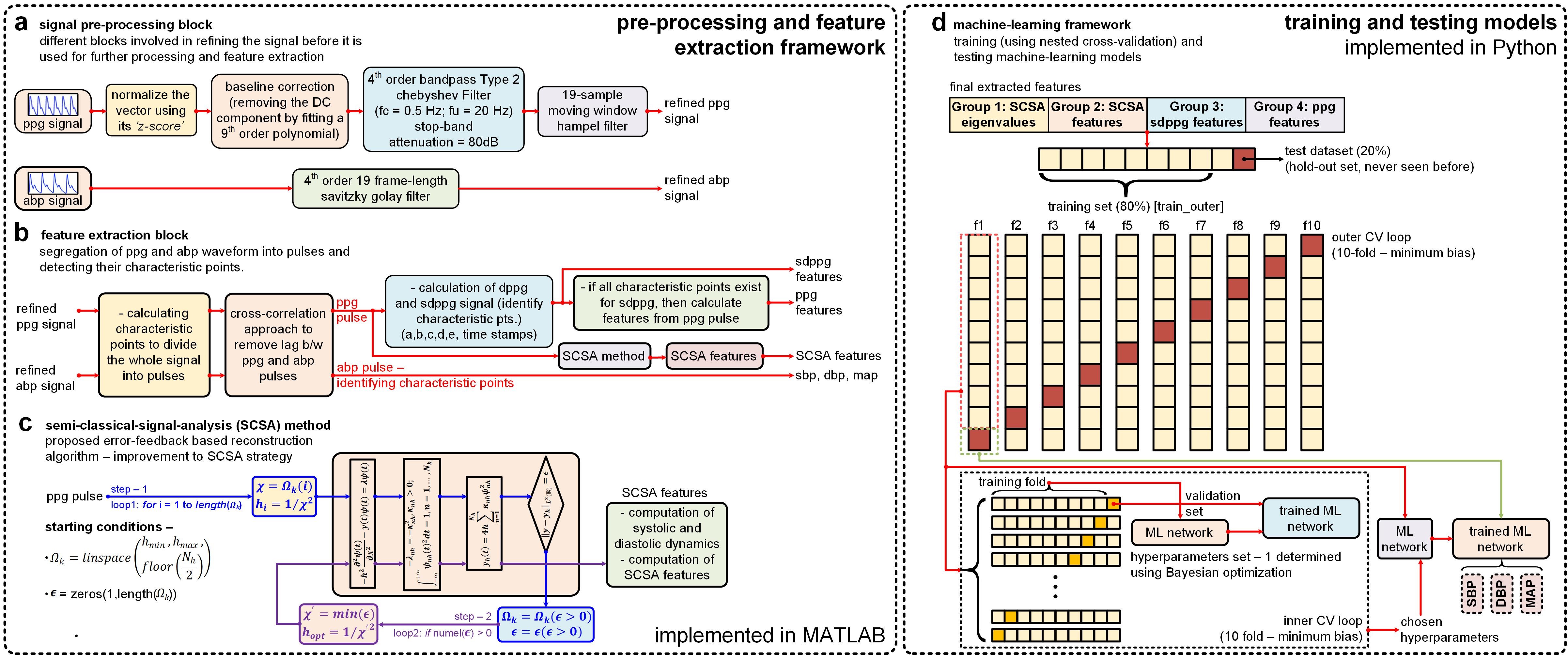}
    \caption{Flow Diagram of the proposed cuffless BP estimation methodology. (a) Signal processing block (b) Feature extraction block, collecting clinical features from PPG and SDPPG signal. (c) Proposed Error Feedback based reconstruction strategy: An improvement to the original SCSA algorithm. (d) Cuffless BP estimation strategy using Machine Learning strategy and models.}
    \label{figure-4}
\end{figure*}
The parameter $\chi$ is used to optimize $h$. Expanding (8) results in $4h\kappa_{1h}\psi_{1h}^2, 4h\kappa_{2h}\psi_{2h}^2,\cdots, 4h\kappa_{N_h h}\psi_{N_h h}^2$, which are termed as the Schr\"odinger components. These components are a series of smooth pulse shaped orthogonal basis functions, also referred to as solitons. These basis functions are localized and selectively capture the Spatio-temporal pattern of $y(t)$. For a large value of $h$, fewer solitons are formed in the series expansion of (8), which fails to capture the precise details of $y(t)$, implementing a bad reconstruction. In contrast, a higher $h$-value implies more components (high value of $N_h$), facilitating high-fidelity reconstruction. As the number of components $(N_h)$ increases, the system becomes computationally expensive to simulate. Therefore, there lies a trade-off in choosing the optimal value of $h$. In this study, the authors propose a novel strategy to solve this trade-off by determining the optimized $h$-value by keeping the computational cost in check. The said algorithm involves a two-step approach. The first step (refer Fig. 4(c), highlighted in blue) involves a loop that iterates over a fixed range of $\chi$ and reconstructs the input pulse at every step. Moreover, the reconstruction error $\epsilon = y - y_h$ is calculated at each step and stored in an array. The second step (Fig. 4(c), highlighted in purple) begins after that by finding the minimum $\epsilon$ and using that to optimize $h$, thereby performing the optimal reconstruction. It was observed that at least $20$ level decomposition $(N_h=20)$ of PPG pulse was necessary to fully capture all the morphological variation of PPG (refer Fig. 2(c)), thereby achieving the least reconstruction error.  

\vspace{-2mm}
\subsection{Feature Extraction}
Establishing a reliable BP estimation model requires identifying features relevant to the task, particularly those that best describe how BP fluctuates. The study involves 38 individual features calculated from the PPG and SDPPG signals to meet these criteria. The definitions of all the features used are summarised in Table I. 
\vspace{3 mm}
\subsubsection{SCSA Features}
According to the SCSA theory, the PPG signals $y$ can be seen as multiple solitons (Schr\"odinger components) interacting in the time $(t)$ dimension. These components can be further segregated into two specific parts. The first part is related to the contraction of the heart or systole, while the second is related to cardiac expansion or diastole. Mathematically, they can be expressed as the decomposition of the PPG pulse into two partial sums, as shown in (9). The first sum runs till $N_s$ largest $\kappa_{nh}$ components, and the second sum is composed of the remaining $\kappa_{nh}$ components. Here, $N_s$ is defined as $1, 2, \cdots, min⁡(3, \sfrac{N_h}{2})$. The first partial sum reflects the rapid phenomena of the systolic phase, while the second partial sum reflects the slow dynamics of the diastole. For this study, $\mathcal{P}_s$ and $\mathcal{P}_d$ are used to denote the systolic and diastolic dynamics, as illustrated in Figs. 3(a, b). [50–52].
\begin{equation}
   \mathcal{P}_s = 4h\sum_{n=1}^{N_s}\kappa_{nh}\psi_{nh}^2, \quad \mathcal{P}_d = 4h\sum_{n = N_s+1}^{N_h}\kappa_{nh}\psi_{nh}^2 
\end{equation}
Several SCSA features are extracted from the pulse and the corresponding systolic and diastolic dynamics [51] such as the decomposed eigenvalues, systolic and diastolic invariant parameters, and the sum of negative eigenvalues. In general, the shape (such as an increase in amplitude and decrease in width) of the PPG signal [48–52] gets modified during their propagation through the arterial network, which is commonly referred to as the ``peaking" and ``steepening" phenomena [52]. The authors propose an additional feature, $PPG_{PSI}$, to quantify this behaviour. It is based on the temporal difference between the maximum and minimum solitons of the systolic and diastolic phases. 
\begin{table*}
    \caption{DEFINITIONS OF THE ANALYSED FEATURES ($N=38$ Features)}
    \centering
    \setlength\tabcolsep{2 pt}
    \setlength{\extrarowheight}{2pt}
    \begin{tabular}{l| c|l}
    \hline\hline
    Group Indices & Expressions & Descriptions \\
    \hline
    & $\kappa_{1h}, \kappa_{2h}, \cdots, \kappa_{20h}$ & Decomposed SCSA negative eigenvalues $(N_h = 20)$ \\
    & $4h\sum_{n=1}^{N_h}\kappa_{nh}$ & Summation of negative eigenvalues $(N_h = 20)$ \\
    SCSA & $4h\sum_{n=1}^{N_s}\kappa_{nh}$ & SCSA Systolic Invariant–I: expressing momentum of the Systolic phase in linear dependence of $\kappa_{nh}$ \\
    FEATURES & $\sfrac{16h}{3}\sum_{n=1}^{N_s}\kappa_{nh}^3$ & SCSA Systolic Invariant–II: expressing momentum in a higher dimension of $\kappa_{nh}$ \\
    & $4h\sum_{n=N_s+1}^{N_h}\kappa_{nh}$ & SCSA Diastolic Invariant–I: expressing momentum of the Diastolic phase in linear dependence of $\kappa_{nh}$ \\
    & $\sfrac{16h}{3}\sum_{n=N_s+1}^{N_h}\kappa_{nh}^3$ & SCSA Diastolic Invariant–II: expressing momentum in a higher dimension of $\kappa_{nh}$ \\
    & $PPG_{PSI}$ & Peaking and Steepening Phenomenon \\
    \hline
    & $\sfrac{b}{a}$ & Ratio of SDPPG b-wave amplitude to a-wave amplitude \\
    & $\sfrac{c}{a}$ & Ratio of SDPPG c-wave amplitude to a-wave amplitude \\
    & $\sfrac{d}{a}$ & Ratio of SDPPG d-wave amplitude to a-wave amplitude \\
    & $\sfrac{e}{a}$ & Ratio of SDPPG e-wave amplitude to a-wave amplitude \\
    SDPPG & $T_a$ & Time delay between a-wave peak and onset point of SDPPG \\
    FEATURES & $T_{ba}$ & Time delay between b-wave peak and a-wave peak of SDPPG \\
    & $T_{cb}$ & Time delay between c-wave peak and b-wave peak of SDPPG \\
    & $T_{dc}$ & Time delay between d-wave peak and c-wave peak of SDPPG \\
    & $T_{ed}$ & Time delay between e-wave peak and d-wave peak of SDPPG \\
    & $AI = \sfrac{(b-c-d-e)}{a}$ & Ratio of a-wave amplitude to the difference of the amplitudes of b, c, d and e-wave \\
    \hline
    PPG & $BW_{66}$ & Branch Width Ratio at 66\% of the pulse height of PPG \\
    FEATURES & $PIR_p$ & Ratio of PPG peak point amplitude to valley point amplitude \\
    \hline\hline
    \end{tabular}
    \label{table-1}
\end{table*}
\vspace{3 mm}
\subsubsection{PPG-SDPPG Features}
Several studies [2–35] have used various PPG and SDPPG-based features for BP estimation. This study collects a total of 12 features from both signals that portray a strong correlation with BP values. The first ten features are extracted from the SDPPG signal, including the Ageing index ($AI$), Amplitude ratios, and the timestamp difference between different SDPPG peaks. The Amplitude ratios $(\sfrac{b}{a}$, $\sfrac{c}{a}$, $\sfrac{d}{a}$, and $\sfrac{e}{a})$ reflect arterial stiffness and the distensibility of the peripheral artery [19]. On the other hand, the timestamps of the characteristic points provide information about different activities during a heartbeat. The last two features are extracted from the PPG signal: (a) Branch width at $66\%$ of the height of the pulse, which is related to total peripheral resistance [36]; and (b) $PIR_p$ [18, 36], linked to changes in arterial diameter. The features are visually described in Figs. 2(a), and (b).

\vspace{-2 mm}
\subsection{Machine Learning Algorithms}
In this study, the authors evaluated the extracted features' performance and their combinations using regression models [7–20]. Four regression (machine learning) models were employed for blood pressure prediction, namely, SVR [11], Category Boosting (CatBoost) [60], Extreme Gradient Boosting (XGBoost) [15, 59], and Light Gradient Boosting (LightGBM) [61]. The first three regression algorithms are widely used and are well-known [15], so the details are not discussed in this manuscript. LightGBM, on the other hand, has not been previously introduced to this field. The ``Light" in the name refers to the computational efficiency of the algorithm, which is much faster than other gradient-boosting algorithms [61]. LightGBM is based on a gradient-boosting structure and uses a tree-based learning algorithm in which the tree grows vertically and horizontally [61]. The algorithm selects the leaf with the maximum delta loss for growth and can reduce the loss further by expanding a similar leaf. It can also handle large amounts of data and requires less memory. All four regression models were trained on the training dataset comprised of the feature vectors listed in Table I. Separate testing data was used to assess the performance of these algorithms. For the rest of the papers, the models with the best performance will be reflected using numbers (performance metrics).

\section{RESULTS AND DISCUSSION}
\subsection {Training and Testing the BP estimation model}
In this study, the authors have applied ML models, including SVR, CatBoost, XGBoost, and LightGBM, to predict BP values using the features listed in Table I. Before starting the training phase, the feature set was normalised by setting the mean to zero and the variance to one. $k$-fold cross-validation (CV) approach was used to divide the normalised feature set into training and testing sets. A nested CV approach was then used for training and selecting the best set of hyperparameters for each model. The value of $k$ was chosen to be 10 because it resulted in lower bias and modest variance in estimates [31, 37, 62]. The training process involved two loops: an inner loop and an outer loop. During the outer loop, the data set was divided into $10$ folds, with nine folds used for training (referred to as \texttt{train\_outer}) and the remaining fold (\texttt{test}) serving as a hold-out set representing unseen data, which was used only for testing the trained model. The \texttt{train\_outer} set was then passed to the inner loop, where it was divided into 9 folds (\texttt{train\_inner}) for training and hyperparameter tuning and one fold (validation set) for validation. For each iteration in the inner loop, the hyperparameters associated with the algorithm were identified on the \texttt{train\_inner} subset using the Bayesian optimisation approach, and the mean absolute error (MAE) was calculated on the validation set. After the iterations in the inner loop, the instance with the lowest MAE and associated hyperparameters was selected as the best choice for the model. Then, in the outer loop, the model instance with the best hyperparameter set was trained on the \texttt{train\_outer} set and tested on the test set. In each fold of the outer CV, the model's hyperparameters were tuned independently to minimise an inner CV estimate of the generalisation performance. The outer loop was essentially estimating the performance of a method for fitting a model. The nested CV approach is more advantageous than normal cross-validation because it reduces the optimisation bias [62]. The best set of hyperparameters for each model is reported in Table II, along with the algorithm’s computational complexity. 

The accuracy of a BP estimation model is typically evaluated using $MAE \pm SDAE$ (mean absolute error $\pm$ standard deviation of absolute errors) or $MAE \pm SD$ (mean absolute error $\pm$ standard deviation) [11–26]. Table III compares the performance of multiple regression algorithms based on MAE, SDAE, and correlation coefficient ($r$). CatBoost [60] and LightGBM [61] performed the best among the regression frameworks. They outperformed traditional algorithms such as SVR and XGBoost. The correlation coefficient ($r$) between true and estimated values in each of the BP criteria (SBP, DBP, and MAP) for both CatBoost and LightGBM are $0.89$ and $0.86$, $0.85$ and $0.82$, and $0.86$ and $0.83$, respectively. These metrics indicate strong regression performance while making CatBoost the top performer. The histograms in Fig. 5(a), (b), and (c) show the distribution of errors for the CatBoost model across all three BP criteria, with errors appearing to be distributed around zero and having a pseudo-normal distribution. The Bland-Altman plots in Fig. 5(d), (e), and (f) present the mean error $(M)$ between the actual and estimated BP values for the CatBoost model. The $M$ values are concentrated around `$0$' and do not deviate significantly, indicating the absence of a fixed bias. Additionally, most data falls within the $M\pm1.96 SD$ range showing no proportional bias between the actual and estimated BP values. For external validation, regression testing was performed using the in-silico Pulse Wave Database (PWD) [55]. $38$ features were extracted from each of the normalised PPG pulses in the PWD, along with their corresponding ABP values, to ensure synchronisation. The plots of the estimated versus real values for systolic, diastolic, and mean absolute BP are shown in Fig. 5(g), (h), and (i). The correlation coefficient $(r)$ between the extracted features and BP values was $0.833$ for SBP, $0.703$ for DBP, and $0.751$ for MAP, indicating a fair degree of linearity. The standard deviation of errors was $6.06$ for SBP, $5.05$ for DBP, and $5.589$ for MAP estimation.

\begin{table*}
    \caption{HYPERPARAMETER TUNING - PARAMETER CHOICES}
    \centering
    \begin{tabular}{l| c|c|c|c|c}
    \hline\hline
    ML Algorithms & Complexity & Associated & \multicolumn{3}{c}{Swept range for various Experiments} \\
    \cline{4-6}
    (Regression) & (Big $\mathcal{O}$ Notation) & Hyperparameters & Exp. - $1^{\mathcal{*}}$ & Exp. - $2^{\mathcal{z}}$ & Exp. - $3^{\mathcal{x}}$ \\
    \hline
    \multirow{5}{*}{SVR} & $\boldsymbol{\mathcal{O}(n^2p + n^3)}$ & Kernel & rbf & - & - \\
                         & $\boldsymbol{p}$ - No. of & Reg. parameter $\mathcal{C}$ & $10,100,\cdots,1000$ & - & - \\
                         & Features & Epsilon $\varepsilon$ & $0.0001,\cdots,0.1$ & - & - \\
                         & $\boldsymbol{n}$ - No. of & Gamma $\gamma$ & `scale', `auto' & - & - \\
                         & Training Sample & Tolerance Criteria & $0.0001,\cdots,0.1$ & - & - \\
    \hline
    \multirow{7}{*}{CatBoost} & $\boldsymbol{\mathcal{O}(npn_{trees})}$ & depth & $6,7,8,\cdots,12$ & $4,6,8,10$ & $6,8,10,12$ \\
                              & $\boldsymbol{n_{trees}}$ - No. of & min\_data\_in\_leaf & $1,100$ & $1,100$ & $1,50$ \\
                              & trees (for & random strength & $0.1,0.2,\cdots,20$ & $0.1,\cdots,20$ & $0.1,\cdots,20$ \\
                              & methods based & learning rate & $0.01,\cdots,0.1$ & $0.01,\cdots,0.1$ & $0.01,\cdots,0.1$ \\ 
                              & on various trees) & l2\_leaf\_reg & $0.01,\cdots,50$ & $0.01,\cdots,10$ & $0.01,\cdots,10$ \\
                              & $\boldsymbol{n,p}$ - holds the & grow policy & SymmetricTree & Lossguide & Depthwise \\
                              & same meaning & leaf est. method & Newton & Gradient & Gradient \\
    \hline
    \multirow{6}{*}{XGBoost} & $\boldsymbol{\mathcal{O}(n^2pn_{trees})}$ & max leaves & $16,64,256,1024$ & - & - \\ 
                             & & max\_bin & $63,255,1024$ & - & - \\
                             & $\boldsymbol{n, p, n_{trees}}$ & eta & $0.01,\cdots,0.1$ & - & - \\
                             & - holds the same & lambda & $0.01,\cdots,10$ & - & - \\
                             & meaning & grow\_policy & loss\_guided & - & - \\
                             & & min\_child\_weight & hist & - & - \\
    \hline
    \multirow{5}{*}{LightGBM} & $\boldsymbol{\mathcal{O}(np)}$ & num\_leaves & $16,64,256,1024$ & - & $16,64,256$ \\
                              & & max\_bin & $63,255,1024$ & - & $63,255$ \\
                              & $\boldsymbol{n,p}$ - holds the & learning\_rate & $0.01,\cdots,0.1$ & - & $0.01,\cdots,0.1$ \\
                              & same meaning & reg\_lambda & $0.01,\cdots,10$ & - & $0.01\cdots,5$ \\
                              & & min\_data\_in\_leaf & $0,20$ & - & $0,100$ \\
    \hline\hline
    \end{tabular}
    \label{table-2}
\end{table*}

\vspace{-1mm}
\begin{figure*}[ht]
    \centering
    \includegraphics[width = 17.5cm, height = 12 cm]{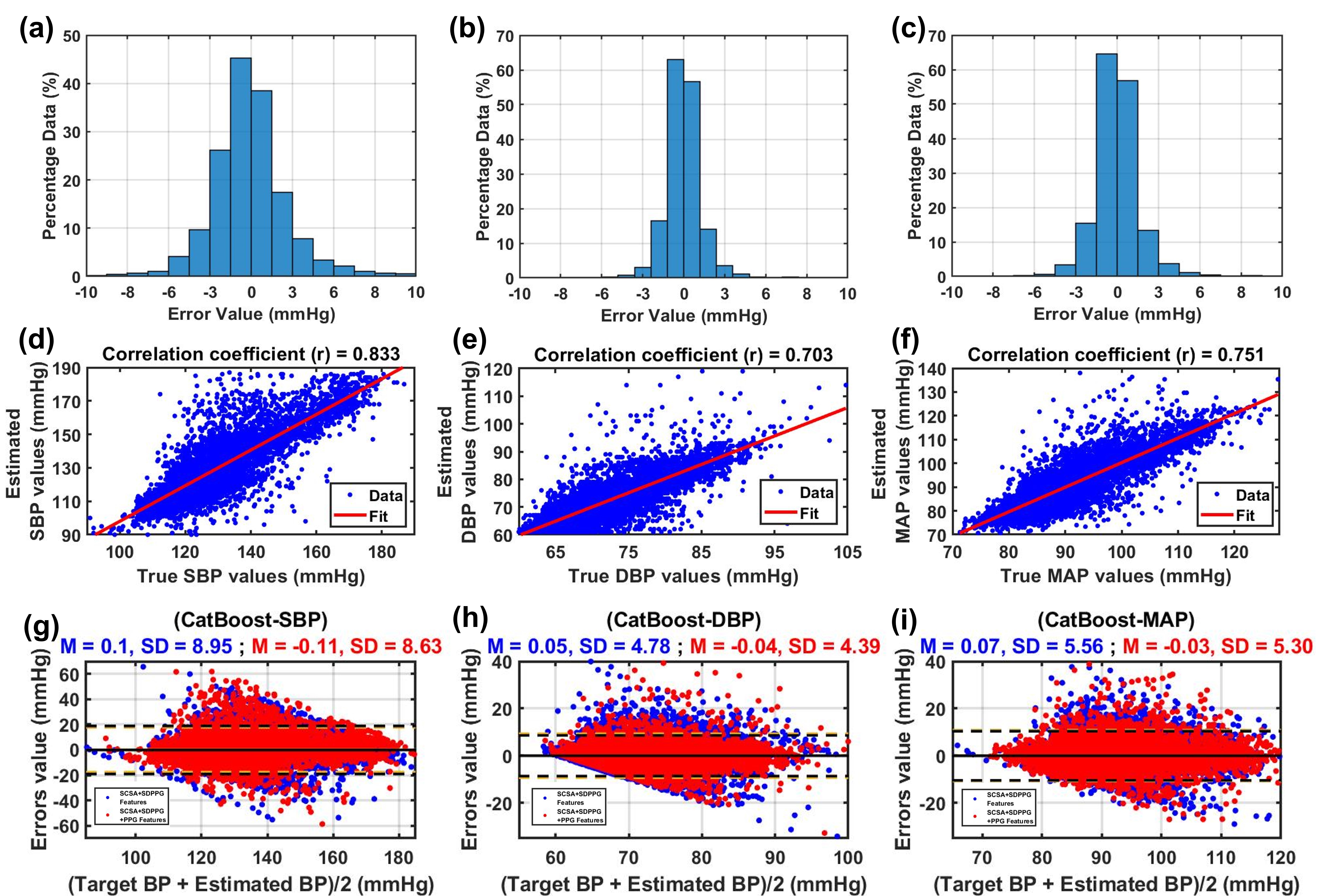}
    \caption{Histograms of errors in the CatBoost model: (a) Systolic Blood Pressure (SBP), (b) Diastolic Blood Pressure (DBP), (c) Mean Absolute Pressure (MAP). Regression plot for (d) SBP, (e) DBP, (f) MAP. Bland–Altman plots of (g) SBP, (h) DBP, and (i) MAP for the CatBoost model. The black lines in the middle show the M, and the dashed lines depict 95\% confidence bounds ($M\pm1.96SD$) for the blue-coloured distribution (SCSA + SDPPG Features). The yellow-coloured lines hold the same meaning as the red-colour distribution (SCSA + SDPPG + PPG Features).}
    \label{figure-5}
\end{figure*}

\begin{table}[ht]
    \caption{ABP ESTIMATION ERROR FOR VARIOUS REGRESSION ALGORITHMS}
    \centering
    \setlength\tabcolsep{3.7 pt}
    \begin{tabular}{c| l|c|c|c|c|c}
    \hline\hline
    BP Criteria & ML & Feature & \multirow{2}{*}{Datasets} & \multicolumn{3}{c}{Evaluation Metrics} \\
    \cline{5-7}
    (mmHg) & Algorithms & Set & & MAE & SDAE & $r$ value \\
    \hline
     \multirow{4}{*}{SBP} & CatBoost & SCSA + & UCI & 5.37 & 5.56 & 0.89 \\
                          & LightGBM & SDPPG + & UCI & 5.49 & 5.92 & 0.86 \\
                          & XGBoost & PPG & UCI & 7.04 & 7.41 & 0.74 \\
                          & SVR & & UCI & 5.96 & 6.45 & 0.81 \\
     \hline 
     \multirow{4}{*}{DBP} & CatBoost & SCSA + & UCI & 2.96 & 3.13 & 0.85 \\
                          & LightGBM & SDPPG + & UCI & 3.29 & 3.47 & 0.82 \\
                          & XGBoost & PPG & UCI & 3.61 & 3.98 & 0.73 \\
                          & SVR & & UCI & 3.24 & 3.61 & 0.80 \\
     \hline 
     \multirow{4}{*}{MAP} & CatBoost & SCSA + & UCI & 3.35 & 3.51 & 0.86 \\
                          & LightGBM & SDPPG + & UCI & 3.77 & 3.81 & 0.83 \\
                          & XGBoost & PPG & UCI & 4.21 & 4.77 & 0.77 \\
                          & SVR & & UCI & 3.89 & 4.32 & 0.80 \\
    \hline\hline
    \end{tabular}
    \label{table-3}
\end{table}

\begin{table}[ht]
    \caption{NORMALIZED FEATURE IMPORTANCE SCORE FOR TOP $15$ FEATURES}
    \centering
    \setlength\tabcolsep{3.2 pt}
    \begin{tabular}{c| c|c|c|c|c|c}
    \hline\hline
    \multirow{2}{*}{Features} & \multicolumn{2}{c|}{Filter Methods} & \multicolumn{3}{c|}{mean(SHAP) values} & Mean \\
    \cline{2-6}
    & $r$ value & MI & XGBoost & CatBoost & LightGBM & Scores \\
    \hline 
    $BW_{66}$ & 0.4231 & 1.0 & 0.5203 & 1.0 & 1.0 & 0.7887 \\
    $T_{cb}$ & 0.9172 & 0.6107 & 0.6169 & 0.8784 & 0.639 & 0.7324 \\
    $\sfrac{c}{a}$ & 1.0 & 0.6698 & 0.4381 & 0.7336 & 0.7668 & 0.7216 \\
    $PIR_p$ & 0.2229 & 0.8612 & 0.498 & 0.8221 & 0.8019 & 0.6412 \\
    $\sfrac{b}{a}$ & 0.6862 & 0.5652 & 0.3357 & 0.7067 & 0.7412 & 0.607 \\
    $PPG_{PSI}$ & 0.0612 & 0.8815 & 0.437 & 0.6691 & 0.6933 & 0.5484 \\
    $\sfrac{d}{a}$ & 0.2262 & 0.6557 & 0.2873 & 0.6411 & 0.7508 & 0.5122 \\
    $T_{dc}$ & 0.2534 & 0.3076 & 0.283 & 0.8733 & 0.8051 & 0.5045 \\
    $\kappa_{6h}$ & 0.3396 & 0.3513 & 1.0 & 0.0387 & 0.0575 & 0.3574 \\
    $AI$ & 0.6427 & 0.4696 & 0.2199 & 0.1657 & 0.2141 & 0.3424 \\
    $T_{ba}$ & 0.4351 & 0.3077 & 0.4331 & 0.2064 & 0.2652 & 0.3295 \\
    $\sfrac{e}{a}$ & 0.255 & 0.3814 & 0.2429 & 0.2304 & 0.3834 & 0.2986 \\
    $T_{ed}$ & 0.3434 & 0.1585 & 0.1478 & 0.3529 & 0.3866 & 0.2778 \\
    $\kappa_{5h}$ & 0.3608 & 0.3263 & 0.4694 & 0.0774 & 0.0863 & 0.264 \\
    $\kappa_{9h}$ & 0.2662 & 0.2651 & 0.6553 & 0.0637 & 0.0831 & 0.2587 \\
    \hline\hline
    \end{tabular}
    \label{table-4}
\end{table}

\vspace{-2 mm}
\subsection{Feature Selection Results}
Feature selection is a process used to identify the most relevant and non-redundant features in a dataset, which can help to improve the understanding of the relationship between individual features and the response variables [16, 31]. In this study, the authors have used six feature selection methods to determine the relative importance of each feature in BP estimation. These methods were divided into filter methods (Pearson Correlation $(r)$ and Mutual Information (MI)) and gradient boosting-based regressors (XGBoost Feature Importance, CatBoost Feature Importance, LightGBM Feature Importance). The feature importance scores calculated using filter methods were normalised to a scale between 0 and 1, and the mean value of the scores for each feature was analysed to identify the best predictors for BP estimation. For gradient boosting-based regressors, we used SHAP values [63]. The results for the top $15$ features are presented in Table IV. It can be observed that PPG \& SDPPG-based features, particularly $BW_{66}$, $PIR_p$, $T_{cb}$, $\sfrac{c}{a}$, and $\sfrac{b}{a}$ are more important than SCSA-based features. This is because PPG-based features have a direct correlation with high BP. Therefore, it is necessary to include PPG and SDPPG features with SCSA-based features to achieve clinical accuracy. 

To further validate the results of the feature selection process, the feature set listed in Table I was divided into three groups ((a) SCSA features, (b) SCSA + SDPPG features, (c) SCSA+ SDPPG+ PPG features), and the MAE and SDAE for each group were calculated. Using only the SCSA features resulted in an $MAE\pm SDAE$ of $9.74\pm10.40$ mmHg for SBP, $4.93\pm5.94$ mmHg for DBP, and $5.92\pm6.02$ mmHg for MAP. Adding the SDPPG features to the SCSA dataset reduced the MAE to $5.37$ mmHg for SBP, $2.96$ mmHg for DBP, and $3.35$ mmHg for MAP. Adding PPG features to the existing dataset further improved the estimation accuracy, as shown in Table III. The authors believe that this happens due to the higher degree of correlation between ABP and morphologically and clinically relevant PPG features. The trend of MAE and SDAE after the addition of consecutive feature groups follows Fig. 6 (a) and (b). 

\vspace{-2 mm}
\subsection{Evaluation considering BP Measurement Standards}
The performance of an ABP estimation method is often evaluated using two standards: AAMI-SP10 and BHS. The AAMI-SP10 standard, set by the Advancements of Medical Instrumentation, states that the mean error $(M)$ and standard deviation of errors $(SD)$ for a BP estimation model should not exceed $5$ and $8$ mmHg, respectively, while considering a mercuric sphygmomanometer based measurement as a reference. The number of subjects $(N)$ in the study should be at least 85 [64] to ensure the generalisation capability of the estimation method. Table V compares the results obtained to the AAMI standards. The $M$ values for all the BP criteria’s using the CatBoost algorithm are within the AAMI limits. However, the $SD$ for SBP estimation slightly exceeds the AAMI requirement, while the SD values for DBP and MAP are within the standard limits. The BHS standard [65] requires that an estimation method achieve an MAE of $\leq5$ mmHg for 60\% of the test dataset, $\leq10$ mmHg for 85\%, and $\leq15$ mmHg for 95\% to be considered Grade A. Table V also compares the results to the BHS standard. Based on this table, the proposed method using the CatBoost algorithm achieves Grade A for all three BP categories, while the LightGBM algorithm achieves Grade A for DBP and MAP and Grade B for SBP.

\vspace{-3.5 mm}
\begin{figure*}[ht]
\includegraphics[width = 16cm, height= 6.3cm]{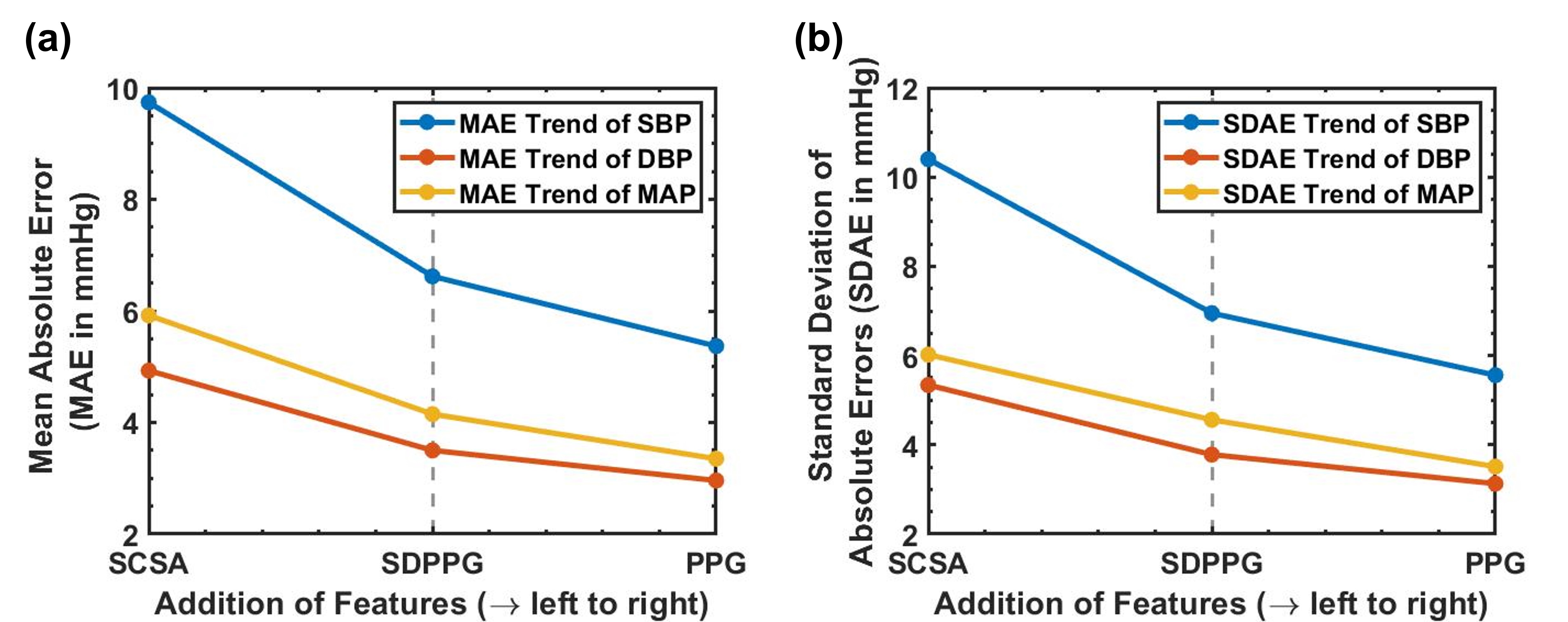}
\centering
\caption{The trend of (a) Mean Absolute Error (MAE) and (b) Standard Deviation of Absolute Errors (SDAE) in SBP, DBP and MAP estimation versus Added Features. New feature groups (Table I) are added to the previous ones from left to right}
\label{figure-6}
\end{figure*}

\begin{table}[ht]
\caption{PERFORMANCE AGAINST AAMI \& BHS STANDARD \\ (Errors in mmHg)}
    \centering
    \setlength\tabcolsep{3.6pt}
    \begin{tabular}{l| c|c|c|c|c|c|c}
    \hline\hline
    ML & BP Criteria & \multicolumn{2}{c|}{Errors} & \multirow{2}{*}{$(N)$} & \multicolumn{3}{c}{Cumulative Error \%} \\
    \cline{3-4}\cline{6-8}
    Algorithms & (mmHg) & $M$ & $SD$ & & $\leq 5$ & $\leq 10$ & $\leq 15$ \\
    \hline
    & SBP & -0.11 & 8.63 & 942 & 61\% & 85\% & 96\% \\
    CatBoost & DBP & -0.04 & 4.39 & 942 & 83\% & 97\% & 100\% \\
    & MAP & -0.03 & 5.30 & 942 & 76\% & 94\% & 99\% \\
    \hline 
    & SBP & -0.09 & 9.11 & 942 & 52\% & 79\% & 90\% \\
    LightGBM & DBP & -0.01 & 4.86 & 942 & 80\% & 96\% & 99\% \\
    & MAP & -0.07 & 5.66 & 942 & 72\% & 93\% & 98\% \\
    \hline 
    \multicolumn{8}{c}{Clinical Blood Pressure Estimation Standards} \\
    \hline 
    \multicolumn{2}{c}{AAMI-SP10 [64]} \vline & $\leq5$ & $\leq8$ & $\geq85$ & - & - & - \\
    \hline 
    \multirow{3}{*}{BHS [65]} & Grade A & - & - & - & 60\% & 85\% & 95\% \\
                              & Grade B & - & - & - & 50\% & 75\% & 90\% \\
                              & Grade C & - & - & - & 40\% & 65\% & 85\% \\
    \hline\hline 
    \end{tabular}
    \label{table-5}
\end{table}

\subsection{Comparison with Other Literature}
Table VI compares the results obtained in this study with previous works that use only PPG signals for estimation, as outlined in Sections 2.2 and 2.3. Comparing related work in this field is challenging due to the limited availability of published Pulse Wave Decomposition analysis results. Hence, we included papers that used PWA and SDPPG features in comparison studies. In most cases, the proposed method outperformed other published results. Most authors failed to indicate whether any data from the test subjects were included in the training data [16]. Hence there is no justification for their obtained results. The lowest errors were achieved on small, selected public or privately collected data subsets. Moreover, the recent growth in the use of Deep Learning algorithms has also contributed to lower estimation errors. But those studies face the common disadvantage of the increased computational burden [73, 74, 76].

\begin{table*}[ht]
\caption{COMPARATIVE ANALYSIS OF PROPOSED ESTIMATION STRATEGY WITH OTHER PUBLISHED WORKS}
    \centering
    \begin{tabular}{l| c|c|c|c|c|c|c}
    \hline\hline
     \multirow{2}{*}{Authors} & \multirow{2}{*}{Features} & Regression & Dataset & \multicolumn{2}{c|}{SBP} & \multicolumn{2}{c}{DBP} \\
    \cline{5-8}
    & & Methods & (Subjects/Signals) & MAE & SDAE & MAE & SDAE \\
    \hline 
    \multirow{2}{*}{Liu [19]} & 14 SDPPG, & \multirow{2}{*}{SVR} & MIMIC II & \multirow{2}{*}{8.54} & \multirow{2}{*}{10.90} & \multirow{2}{*}{4.34} & \multirow{2}{*}{5.80} \\
    & 21 PPG & & (910 PPG cycles) & & & & \\
    \hline 
    \multirow{2}{*}{Kachuee [11]} & PAT, HR & \multirow{2}{*}{AdaBoost} & MIMIC II & \multirow{2}{*}{11.17} & \multirow{2}{*}{10.09} & \multirow{2}{*}{5.35} & \multirow{2}{*}{6.14} \\
    & time intervals & & (5559 PPG signals) & & & & \\
    \hline 
    \multirow{2}{*}{Zhang [66]} & 21 Features, & \multirow{2}{*}{SVR} & Queensland & \multirow{2}{*}{11.64} & \multirow{2}{*}{8.20} & \multirow{2}{*}{7.62} & \multirow{2}{*}{6.78} \\
    & Physiological & & (7000 PPG cycles) & & & & \\
    \hline 
    \multirow{2}{*}{Slapni\v car [75]} & Frequency, time & Regression & MIMIC & \multirow{2}{*}{7.87} & \multirow{2}{*}{7.47} & \multirow{2}{*}{3.84} & \multirow{2}{*}{3.63} \\
    & Physiological & Tree & (41 Patients) & & & & \\
    \hline 
    \multirow{2}{*}{Mousavi [67]} & Features from & \multirow{2}{*}{SVR, RF} & MIMIC II & \multirow{2}{*}{3.97} & \multirow{2}{*}{8.90} & \multirow{2}{*}{2.43} & \multirow{2}{*}{4.17} \\
    & PPG, SDPPG & & (1323 PPG signals) & & & & \\
    \hline 
    \multirow{2}{*}{Laleg [46]} & (SCSA, PAT) & \multirow{2}{*}{SVR} & MIMIC II & \multirow{2}{*}{6.95} & \multirow{2}{*}{6.89} & \multirow{2}{*}{4.88} & \multirow{2}{*}{5.54} \\
    & from PPG & & (6760 PPG signals) & & & & \\
    \hline
    \multirow{2}{*}{Moajjem [31]} & 101 PPG & Gaussian & PPG-BP-Database & \multirow{2}{*}{3.02} & \multirow{2}{*}{9.29} & \multirow{2}{*}{1.74} & \multirow{2}{*}{5.54} \\
    & (time \& frequency) & Process & (219 Patients) & & & & \\  
    \hline 
    \multirow{2}{*}{Navid [16]} & 19 PPG & \multirow{2}{*}{AdaBoost} & MIMIC II & \multirow{2}{*}{8.22} & \multirow{2}{*}{10.38} & \multirow{2}{*}{4.17} & \multirow{2}{*}{4.22} \\
    & Features & & (12k PPG signals) & & & & \\
    \hline
    \multirow{2}{*}{Yang [36]} & 28 PPG & LR, RF & VitalDB Database & \multirow{2}{*}{5.07} & \multirow{2}{*}{6.92} & \multirow{2}{*}{2.86} & \multirow{2}{*}{3.99} \\
    & Features & ANN, RNN & (1376 Patients) & & & & \\
    \hline 
    \multirow{2}{*}{El-Hajj [68]} & 59 PPG & Bi-GRU/ & MIMIC II & \multirow{2}{*}{1.26} & \multirow{2}{*}{1.63} & \multirow{2}{*}{2.58} & \multirow{2}{*}{3.35} \\
    & Features & Attention & (500 PPG signals) & & & & \\
    \hline 
    \multirow{2}{*}{Lee [69]} & 7 PPG & \multirow{2}{*}{Bi-LSTM} & Collected & \multirow{2}{*}{5.82} & \multirow{2}{*}{6.82} & \multirow{2}{*}{5.24} & \multirow{2}{*}{6.06} \\
    & Features & & (18 Patients) & & & & \\
    \hline 
    \multirow{2}{*}{This work} & `38': SCSA, & \multirow{2}{*}{CatBoost} & MIMIC II & \multirow{2}{*}{5.37} & \multirow{2}{*}{5.56} & \multirow{2}{*}{2.96} & \multirow{2}{*}{3.13} \\
    & SDPPG, PPG & & (12j PPG signals) & & & & \\
    \hline\hline
    \end{tabular}
    \label{table-6}
\end{table*}

\vspace{-1.5 mm}
\subsection{Performance in Clinically Relevant Database}
\begin{figure*}[ht]
    \centering
    \includegraphics[width = 18cm, height = 4.0 cm]{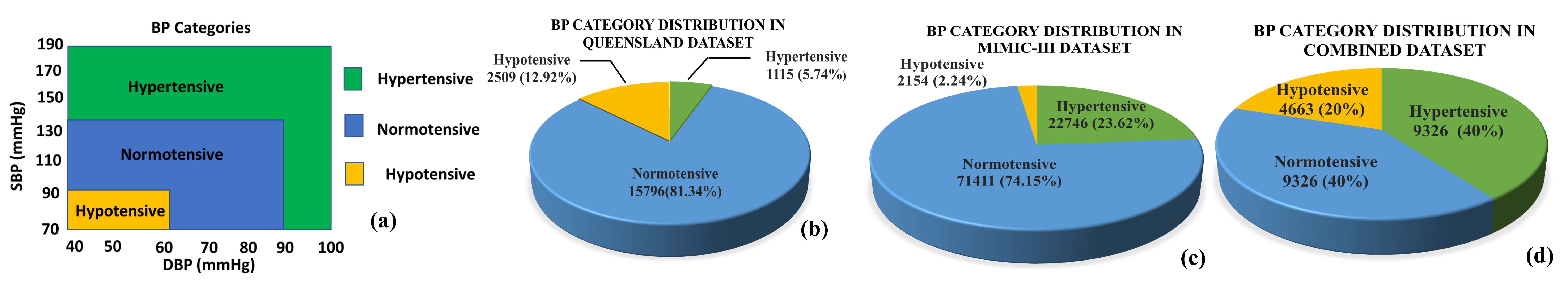}
    \caption{(a) BP classification scheme and distribution for different BP categories. Reference BP categorical distribution in individual datasets (b) Queensland, (c) MIMIC–III (34 subparts), (d) Combined constructed database.}
    \label{figure-7}
\end{figure*}

Many published works have not evaluated BP estimation accuracy separately for different clinical BP categories (hypotensive, normotensive, and hypertensive), as shown in Fig. 7(a). Hosanee \textit{et al.} [4] found that only 8\% of analysed studies included hypertensive patients, while 52\% used normotensive patients. Accurate evaluation of hypotensive and hypertensive BP is essential for the early diagnosis of many diseases [3, 16]. The high accuracy for normotensive BP estimation may explain why most current cuffless BP estimation algorithms do not perform well in hypotensive or hypertensive datasets [16, 58]. In [58], category-specific BP algorithms were developed, but this method is not suitable for real-time BP estimation due to its increased complexity. Hasanzadeh \textit{et al.} [16] used the UCI database and divided it into subsets of low and high BP values for both systolic and diastolic criteria, but their proposed estimation method performed poorly. The UCI dataset contains a minimal number of hypertensive patients [58]. Hence, clinical BP category-based evaluation on that dataset will not suffice our objective. Therefore, to evaluate the proposed framework’s efficacy in clinical BP categories, an additional database using two publicly available datasets (the Queensland Dataset, and the MIMIC-III Dataset) was created. The Queensland Dataset contains PPG and ABP records from 32 subjects over 13 minutes to 5 hours (median 105 minutes) [57]. However, it has a relatively small number of hypotensive and hypertensive BP values, as shown in Fig. 7(b). We extracted 5-second segments from each PPG record and applied signal quality assessment to remove poor segments [53]. This resulted in the following BP categories in the Queensland Dataset: Hypotensive (2509 segments, 12.92\%), normotensive (15796 segments, 81.34\%), and hypertensive (1115 segments, 5.74\%) segments. The MIMIC-III Dataset [56] contains physiological measurement data with a significant number of normotensive and hypertensive categories but has a limited number of hypotensive BP values, as shown in Fig. 7(c). This study includes PPG and ABP data from 720 individuals in the MIMIC-III dataset and applies the same signal quality assessment [53] to segment out the poor signals. This contributed to the following percentage of data from MIMIC-III: hypotensive (2154 segments, 2.24\%), normotensive (71411 segments, 74.15\%), and hypertensive (22746 segments, 23.62\%). However, combining the two datasets directly would create imbalances in the distribution of BP categories. Therefore, the authors treat the hypotensive data as the benchmark by combining 2509 segments from the Queensland Dataset and 2154 segments from the MIMIC-III Dataset while assuming it as 20\% of the combined signal segments. Similarly, in the normotensive (40\%) and hypertensive (40\%) categories, signal segments are selected from Queensland and MIMIC-III datasets in equal proportions. The final database distribution is shown in Fig. 7(d). The CatBoost algorithm was employed to estimate BP on the constructed database. When applied to the Queensland Dataset (combining all BP categories), CatBoost provided the following results for SBP estimation: MAE (ME) of 8.43 (0.23) mmHg and SDAE (SDE) of 7.8 (8.86) mmHg. For DBP, CatBoost had an MAE (ME) of 6.27 (0.18) mmHg and SDAE (SDE) of 5.78 (6.91) mmHg. On the MIMIC-III dataset, CatBoost achieved an MAE (ME) of 8.06 (0.32) mmHg and SDAE (SDE) of 8.15 (9.07) mmHg for SBP, and an MAE (ME) of 3.96 (0.15) mmHg and SDAE (SDE) of 3.92 (4.35) mmHg for DBP. For both datasets, the CatBoost algorithm met the BHS criteria with a grade A for all three BP criteria. The performance of the combined database in the three different BP categories is shown in Table VII. It can be concluded that the proposed algorithm performs well in the hypotensive and hypertensive categories but slightly out of the AAMI-SP10 and BHS thresholds in the normotensive category for SBP.

\begin{table*}[ht]
\caption{ESTIMATION ACCURACY IN CONSTRUCTED DATABASE}
    \centering
    \setlength\tabcolsep{3.7 pt}
    \setlength{\extrarowheight}{0.5pt}
    \begin{tabular}{l|c| c|c|c|c|c|c|c|c}
    \hline\hline
    BP Category & BP Criteria & \multicolumn{5}{c|}{Evaluation Metrics} & \multicolumn{3}{c}{Cumulative Error Percentage}\\
    \cline{3-10}
    (mmHg) & (mmHg) & MAE & SDAE & M & SD & $r$ & $\leq5$ mmHg & $\leq10$ mmHg & $\leq15$ mmHg \\
    \hline
    \multirow{3}{*}{Hypotensive} & SBP & 2.06 & 2.65 & 0.008 & 2.85 & 0.88 & 91\% & 100\% & 100\% \\
    & DBP & 1.73 & 2.18 & 0.10 & 2.38 & 0.91 &95\% &100\% & 100\% \\ 
    & MAP & 1.62 & 2.02 & -0.04 & 2.20 & 0.92 &96\% &100\% &100\%  \\
    \hline 
    \multirow{3}{*}{Normotensive} & SBP & 7.90 & 8.29 & -0.30 & 10.35 & 0.75 & 42\%  & 71\% & 86\% \\ 
    & DBP & 5.66 & 6.57 & 0.15 & 7.65 & 0.81 & 57\% & 84\% & 94\% \\
    & MAP & 4.48 & 5.19 & 0.22 & 7.20 & 0.83 & 57\%  & 85\% & 96\% \\
    \hline 
    \multirow{3}{*}{Hypertensive} & SBP & 5.61 & 6.37 & 0.45 & 7.87 & 0.83 & 60\% &85\% &94\% \\ 
    & DBP & 5.13 & 6.04 & -0.02 & 7.04 & 0.82 & 62\% & 86\% & 96\% \\
    & MAP & 3.76 & 4.93 & 0.07 & 4.93 & 0.86 &72\% &95\% &100\% \\
    \hline\hline
    \end{tabular}
    \label{table-7}
\end{table*}
\begin{figure*}[ht]
    \centering
    \includegraphics[width = 18cm, height = 11.3cm ]{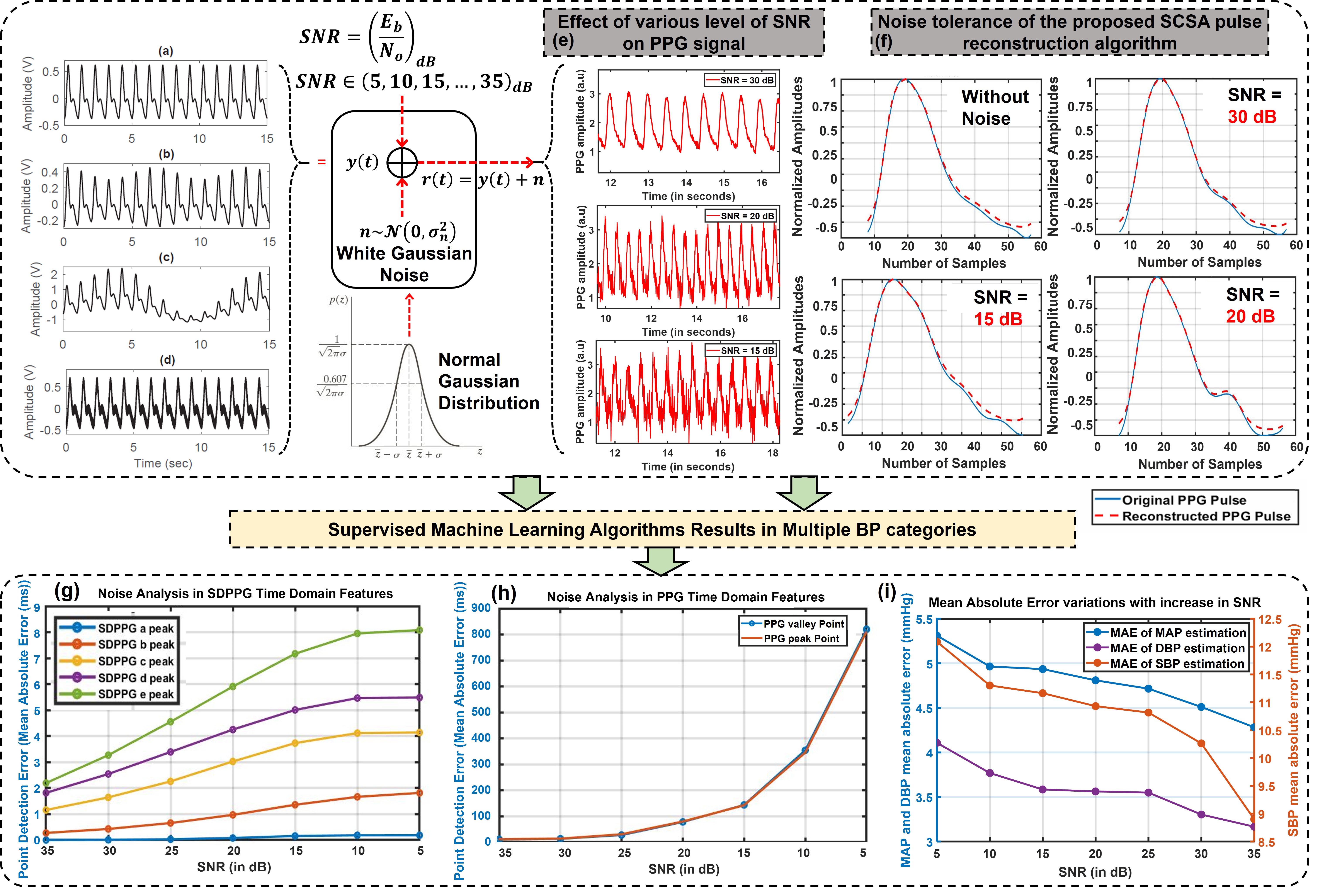}
    \caption{Idealised examples of various scenarios of PPG signals: (a) Clean, (b) Respiratory modulated, (c) Baseline modulated, (d) Motion affected. (e) White Gaussian enhanced signal generation technique and sample signals at different SNR (5,10,15,…,30,35 dB) levels. (f) Signal reconstruction capability test for SCSA algorithm at different SNR levels, explaining the effect of noise on the time domain. Point Detection error of different time-domain (g) SDPPG peak and valley points (h) PPG peak and valley points. (i) MAE of different BP categories at different SNR levels using the CatBoost Algorithm.}
    \label{figure-8}
\end{figure*}
\vspace{-2 mm}
\subsection{Performance in Noisy Database}
The algorithms proposed in previous literature are often developed, calibrated, and validated based on offline datasets such as MIMIC II [11-16], MIMIC III [56], Queensland [57], and Vital DB [36]. However, these datasets have limitations, as the BP of patients is influenced by drugs [16, 36, 48], and the patient’s age in these datasets is generally higher than that of healthy adults [70–72]. Therefore, it is vital to validate the learning algorithms developed for BP estimation on a diverse range of subjects, environments, and motion artefacts before they can be commercialised, as required by IEEE protocols [72]. Testing the algorithm on a non-medical dataset with a range of body postures, movements, and age groups could demonstrate its generalizability for accuracy evaluation. In this study, the authors used a wearable collected database [71] available on ‘IEEE DataPort’ to validate the efficacy of the developed algorithm. The database includes continuous ABP waveform measured via the Portapres device by Finapres Medical Systems, Netherlands, and ECG and PPG measured using the Astroskin wearable body metrics vest by Carré Technologies Inc., Canada [71]. The database was collected from five young, healthy individuals (4 males and 1 female, average age 28 years) with varying fitness levels and no history of cardiovascular or peripheral vascular disease. The subjects engaged in various activities during data collection, such as walking, running, working, cycling, performing Valsalva maneuvers, and performing static handgrip exercises [70–72]. The CatBoost and LightGBM algorithms were used to evaluate the performance of the proposed framework on the Noisy Dataset. Table VIII compares the performance of both regression frameworks. However, assessing the results using the AAMI and BHS standards is unfeasible, as there are not enough subjects in the dataset.

\begin{table}[ht]
    \caption{PERFORMANCE EVALUATION IN NOISY DATASET}
    \centering
    \setlength\tabcolsep{3.2 pt}
    \begin{tabular}{l| c|c|c|c|c}
    \hline\hline
    ML & BP Criteria & \multicolumn{4}{c}{Evaluation Metrics} \\
    \cline{3-6}
    Algorithms & (mmHg) & MAE & SDAE & M & $r$ value \\
   \hline
   \multirow{3}{*}{CatBoost} & SBP & 6.14 & 7.18 & -0.17 & 0.78 \\
                             & DBP & 4.20 & 5.57 & -0.12 & 0.62 \\
                             & MAP & 4.48 & 4.93 & 0.05 & 0.71 \\
   \hline 
   \multirow{3}{*}{LightGBM} & SBP & 6.25 & 7.25 & -0.46 & 0.78 \\
                             & DBP & 4.29 & 5.45 & -0.15 & 0.63 \\
                             & MAP & 4.54 & 5.13 & -0.18 & 0.71 \\
    \hline\hline
    \end{tabular}
    \label{table-8}
\end{table}
\vspace{-2mm}
\subsection {Noise Stress Test}
The authors conducted experiments by adding white Gaussian noise to raw PPG signals at different signal-to-noise ratios (SNR) to further establish the proposed method's performance in the presence of noise, as shown in Fig. 8(a)–(i). Pre-processing, key point detection, and pulse reconstruction were performed using the SCSA algorithm at different SNRs. The results in Fig. 8(f) demonstrate the SCSA algorithm's superior noise tolerance capability. During key point detection, the time displacement between salient points in the noisy and raw PPG signals was used as a performance indicator. Fig. 8(g) and (h) show the calculated time errors in key features at different SNRs. Based on Fig. 8(g), the SDPPG key points have superior noise tolerance compared to PPG key points. This is likely because the second derivative transform of the SDPPG acts as a derivative filter, sharpening the curve and reducing noise and timestamp errors. We evaluated the mean absolute error (MAE) of the three BP categories at different SNRs to understand the effect of noise on the performance of the BP estimation algorithm, as shown in Fig. 8(i). 
The results show that errors at SNRs greater than 10 dB are tolerable, and the proposed algorithm can still accurately calculate BP. Additionally, the authors introduced low-frequency sinusoids (baseline wandering) to raw PPG signals during the baseline wandering study and investigated their impact in the pre-processing step. It was observed that the proposed algorithm could effectively remove the baseline wandering effect.

\section{CONCLUSION}
This paper proposed an ABP estimation method using a modified SCSA technique and ML algorithms by processing the PPG signals’ clinically relevant time-domain features and the SDPPG signal. A novel error feedback-based reconstruction algorithm optimises the ‘semi-classical constant’ value and eliminates the trade-off in reconstruction. The proposed method is validated in a virtual in-silico dataset and multiple publicly available databases containing many subjects. Results demonstrate the robustness and statistical reliability of the proposed method, as it satisfies the AAMI standard requirements and achieves Grade A in the BHS protocol. The efficacy of the SCSA method for BP estimation in a noisy database is validated from consumer wearables' data during synchronised physical exercise. Noise stress tests reveal that key feature detection, signal reconstruction capability and estimation accuracy hold well up to a 10 dB SNR ratio. The proposed algorithm works well even in the clinically relevant dataset developed by combining MIMIC-III and Queensland datasets. BP estimated from the proposed method performs better than most of the ML-based and few DL-based methods and paves the way for wearable realisation for cuffless BP estimation.

%\section*{Data and Code Availability}
% The code used to generate the results reported in this paper will be made available on  \href{https://github.com/aayushmanghosh}{\textcolor{blue}{GitHub-repo}} upon publication of this manuscript and can also be available from the corresponding author upon reasonable request. The data used in this study are publicly available and have been explicitly described in the main text. A link to the intermediate datasets (the feature vectors generated at each step) used in our study will also be published in the same GitHub-repo.
%%%%%%%%%%%%%%%%%%%%%%%%%%%%%%%%%%%%%%%%%%%%%%%%%%%%%%%%%%%%%%%%%%%%%%%%%%%%%%%%

%References are important to the reader; therefore, each citation must be complete and correct. If at all possible, references should be commonly available publications.

%%%%%%%%%%%%%%%%%%%%%%%%%%%%%%%%%%%%%%%%%%%%%%%%%%%%%%%%%%%%%%%%%%%%%%%%%%%%%%%%

%%%%%%%%%%%%%%%%%%%%%%%%%%%%%%%%%%%%%%%%%%%%%%%%%%%%%%%%%%%%%%%%%%%%%%%%%%%%%%%%

%References are important to the reader; therefore, each citation must be complete and correct. If at all possible, references should be commonly available publications.

\end{document}